# Novel Radiomic Measurements of Tumor-Associated Vasculature Morphology on Clinical Imaging as a Biomarker of Treatment Response in Multiple Cancers

Nathaniel Braman,[1,2] Prateek Prasanna,[1,3] Kaustav Bera,[1,4] Mehdi Alilou,[1] Mohammadhadi Khorrami,[1] Patrick Leo,[1] Maryam Etesami,[5] Manasa Vulchi,[6] Paulette Turk,[6] Amit Gupta,[4] Prantesh Jain,[4] Pingfu Fu,[1] Nathan Pennell,[6] Vamsidhar Velcheti,[7] Jame Abraham,[6] Donna Plecha[4] and Anant Madabhushi.[1,8]

[1]Case Western Reserve University, Cleveland, OH
[2]Picture Health, Cleveland, OH
[3]Stony Brook University, New York, NY
[4]University Hospitals Cleveland Medical Center, Cleveland, OH
[5]Yale School of Medicine, Department of Radiology & Biomedical Imaging, New Haven, CT
[6]The Cleveland Clinic Foundation (CCF), Cleveland, OH
[7]New York University (NYU) Langone Medical Center, New York, NY, USA
[8]Louis Stokes Cleveland Veterans Medical Center, Cleveland, OH

**Running title:** Vessel Shape Predicts Chemotherapy Response and Prognosis

**Corresponding Author:**
    Anant Madabhushi
    Department of Biomedical Engineering,
    Case Western Reserve University Wickenden Bldg., Rm. 340
    10900 Euclid Avenue Cleveland, Ohio 44106-7207
    axm788@case.edu
    216.368.8519

**Conflict of interest statement:** Nathaniel Braman is a current employee of Picture Health and a former employee of Tempus Labs and IBM Research, with whom he is an inventor on several pending patents pertaining to medical image analysis. He additionally holds equity in Picture Health and Tempus Labs. Prateek Prasanna is a former employee of GE Global Research. Dr. Madabhushi is a co-founder and equity holder in Picture Health. He is additionally an equity holder in Elucid Bioimaging and in Inspirata Inc. He has served as a scientific advisory board member for Inspirata Inc, Astrazeneca, Bristol Meyers-Squibb and Merck. Currently he serves on the advisory board of Aiforia Inc. He also has sponsored research agreements with Philips, AstraZeneca, Boehringer-Ingelheim and Bristol Meyers-Squibb. His technology has been licensed to Elucid Bioimaging. He is also involved in a NIH U24 grant with PathCore Inc, and 3 different R01 grants with Inspirata Inc. Several of the authors hold patents related to the contents of this manuscript: US Patent 10,861,152, "Vascular Network Organization via Hough Transform (VaNGoGH): A radiomic biomarker for diagnosis and treatment response" (Nathaniel Braman, Anant Madabhushi, Prateek Prasanna); US Patent 10,064,594. "Characterizing Disease and Treatment Response with Quantitative Vessel Tortuosity Radiomics" (Nathaniel Braman, Anant Madabhushi, Mehdi Alilou).

# Abstract

**Purpose:** The tumor-associated vasculature differs from healthy blood vessels by its convolutedness, leakiness, and chaotic architecture, and these attributes facilitate the creation of a treatment resistant tumor microenvironment. Measurable differences in these attributes might also help stratify patients by likely benefit of systemic therapy (e.g. chemotherapy). In this work, we present a new category of computational image-based biomarkers called quantitative tumor-associated vasculature (QuanTAV) features, and demonstrate their ability to predict response and survival across multiple cancer types, imaging modalities, and treatment regimens involving chemotherapy.

**Experimental Design:** We isolated tumor vasculature and extracted mathematical measurements of twistedness and organization from routine pre-treatment radiology (computed tomography or contrast-enhanced MRI) of a total of 558 patients, who received one of four first-line chemotherapy-based therapeutic intervention strategies for breast (n=371) or non-small cell lung cancer (NSCLC, n=187).

**Results:** Across four chemotherapy-based treatment strategies, classifiers of QuanTAV measurements significantly ($p<.05$) predicted response in held out testing cohorts alone (AUC=0.63-0.71) and increased AUC by 0.06-0.12 when added to models of significant clinical variables alone. Similarly, we derived QuanTAV risk scores that were prognostic of recurrence free survival in treatment cohorts who received surgery following chemotherapy for breast cancer (p=0.0022, HR=1.25, 95% CI 1.08-1.44, C-index=.66) and chemoradiation for NSCLC (p=0.039, HR=1.28, 95% CI 1.01-1.62, C-index=0.66). From vessel-based risk scores, we further derived categorical QuanTAV high/low risk groups that were independently prognostic among all treatment groups, including NSCLC patients who received chemotherapy only (p=0.034, HR=2.29, 95% CI 1.07-4.94, C-index=0.62). QuanTAV response and risk scores were independent of clinicopathological risk factors and matched or exceeded models of clinical variables including post-treatment response.

**Conclusions:** Across these domains, we observed an association of vascular morphology on CT and MRI – as captured by metrics of vessel curvature, torsion, and organizational heterogeneity – and treatment outcome. Our findings suggest the potential of shape and structure of the tumor-associated vasculature in developing prognostic and predictive biomarkers for multiple cancers and different treatment strategies.

*Statement of translational relevance*

In this study, we introduced a new class of imaging biomarkers measuring the shape and architecture of the tumor-associated vasculature (TAV). We developed and validated TAV models for prediction and prognostication in multiple cancers (breast and non-small cell lung), imaging modalities (computed tomography and contrast-enhanced MRI), and four therapeutic regimens including chemotherapy. We showed across this array of clinical problems that morphology of the TAV correlated with post-treatment response and prognosis, with chaotically organized vasculature prior to treatment generally portending poor outcome. Unlike many computational approaches for prediction and prognosis from clinical imaging – which largely rely on algorithmically complex "black box" machine learning tools such as deep neural networks or abstracted quantitative measures – our approach is rooted directly in the underlying cancer biology of tumor angiogenesis. Accordingly, it is highly clinically interpretable.

# Introduction

Neoadjuvant chemotherapy, or chemotherapy administered prior to surgical intervention, often constitutes first-line intervention in a number of cancer domains.[1–4] When successful, neoadjuvant chemotherapy can offer substantial benefits for patients by reducing tumor burden and increasing a patient's surgical options.[5] However, many patients ultimately fail to respond and, accordingly, will endure financial burden and dangerous side effects without tangible benefit.[6] Furthermore, in many cancers, including breast (BRCA) and non-small cell lung cancer (NSCLC), there is a current lack of validated predictive and prognostic biomarkers capable of definitively guiding first-line chemotherapeutic interventions.[7–9]

Tumor angiogenesis has long been shown to be crucial in cancer progression. Through influence over the body's machinery for synthesizing vasculature, a tumor will initiate the rapid formation of new blood vessels from preexisting vessels in its surrounding peritumoral environment. This newly formed vessel network, known as the tumor-associated vasculature (TAV), enables tumor growth by perfusing it with abundant oxygen and nutrients, as well as providing an avenue for metastatic spread.[10] Histological and molecular evidence of elevated tumor angiogenesis, such as increased density of micro-vessels measured via immunostaining or elevated vascular endothelial growth factor (VEGF) expression,[11] is associated with poor prognosis and therapeutic response.

However, the TAV also possesses crucial architectural differences from healthy blood vessels that are undetectable through routine clinical assessment. Excessive up-regulation of angiogenesis creates vessels that are twisted, leaky, and chaotically organized.[12–15] Previous work has shown that abnormalities in the shape of tumor vessels are detectable on CT and MRI scans and can distinguish cancer from benign lesions.[16–18] This aberrant vessel morphology has been implicated in potentiating treatment refractoriness by reducing drug transfer to the tumor bed, thus leading to a lack of durable response.[19] Conversely, successful normalization of TAV architecture through anti-angiogenic therapy can promote the efficacy of therapeutic intervention.[20] It is likely that tumors that are resistant to treatment will differ in the twistedness and arrangement of their vasculature relative to responsive tumors,[21] which in turn could potentially be captured quantitatively on radiologic imaging.[22] Consequently, computerized analysis of TAV morphology and spatial organization might enable better guidance of chemotherapy-based treatment by stratifying patients according to likely therapeutic benefit.

In this paper, we present a new computational imaging biomarker based on quantitative tumor-associated vasculature (QuanTAV) measurements to characterize the morphology and architecture of the vessel network surrounding a tumor on radiology scans. We present and evaluate a number of computationally extracted measurements of the twistedness and organization of tumor vessels on pre-treatment contrast-enhanced magnetic resonance imaging (MRI) of breast cancer patients and computed tomography (CT) of lung cancer patients. We further demonstrate the predictive and prognostic utility of QuanTAV measurements in the context of response to chemotherapy-related treatments for four cases involving breast and lung cancers across these modalities. In total, the prognostic and predictive utility of QuanTAV was evaluated on 558 patients, including 242 breast cancer patients receiving anthracycline-based neoadjuvant chemotherapy [BRCA-ACT], 129 breast cancer patients receiving neoadjuvant chemotherapy with HER2-targeted therapy [BRCA-TCHP], 97 NSCLC patients receiving platinum-based chemotherapy without surgery [NSCLC-PLAT], and 90 NSCLC patients receiving a trimodality regimen of neoadjuvant chemoradiation followed by surgical intervention [NSCLC-TRI].

## Materials and Methods

### Overview

From 3-dimensional volumes delineating a tumor and its corresponding vasculature, our approach mathematically characterizes the complexity of the tumor-associated vasculature for use in machine learning models to predict outcomes. Vessel volumes are algorithmically reduced[23] to centerlines and split into discrete branches prior to analysis. A set of 91 QuanTAV measurements are then computed, belonging to one of two categories:

- *QuanTAV morphology*[17] (61 features): Features describing the 3D shape of tumor vessels. Metrics measuring the twistedness of vessels across different length scales are calculated: torsion (twistedness across a full vessel branch) and curvature (local twistedness among adjacent points along a branch). Additional metrics such as vessel volume and length, and the proportion of vessels entering a tumor, are also derived.
- *QuanTAV Spatial Organization*[24] (30 features): Features quantifying the degree of heterogeneity in the architecture of the tumor vasculature. 2D projections of the tumor vasculature are generated across each dimension of the imaging plane and in a spherical coordinate system relative to the tumor centroid within a fixed radius of the tumor. The set of QuanTAV Spatial Organization features are statistics describing vessel orientations across each projection image.

For each treatment group, we derive QuanTAV response and risk scores from these metrics, then evaluate their ability to predict response and time to recurrence or progression. Our experimental workflow is summarized in Figure 1. Code for performing QuanTAV analysis and a workable demo are made available at: https://github.com/ccipd/QuanTAV

### Datasets

This Health Insurance Portability and Accountability Act of 1996 regulations–compliant study was approved by the institutional review boards at the University Hospitals Cleveland Medical Center, Cleveland, Ohio and the Cleveland Clinic Foundation and the need for informed consent was waived.

**Breast:** A total of 470 patients who received breast neoadjuvant chemotherapy with pre-treatment dynamic contrast-enhanced (DCE) MRI were identified for this study. Each breast MRI exam consisted of several T1-weighted acquisitions, including a pre-contrast scan and several scans acquired following the injection of gadolinium-based contrast agent. 31 patients were excluded due to poor image quality resulting in flawed vascular segmentation (including low spatial resolution, insufficient temporal scans or poor temporal resolution, severe artifacts, or inadequate vessel enhancement). 68 patients were HER2-positive, but received treatment prior to the introduction of anti-HER2 agents, and were thus excluded from analysis. The total number of patients for analysis was 371. Patient response was defined as pathologic complete response (pCR) following chemotherapy, the most commonly utilized surrogate endpoint in the breast neoadjuvant chemotherapy setting[25,26] and defined as a lack of remaining invasive cancer cells within the breast or axilla based on pathological examination of excised surgical samples (ypT0/isN0), 115 achieved pCR, while 256 retained the presence of residual disease following chemotherapy (non-pCR). Patients received different chemotherapeutic regimens based on the expression of the HER2 receptor protein, and patients were split into corresponding treatment groups for analysis.

- *BRCA-ACT:* 242 patients were HER2-negative, and received an anthracycline-based regimen with or without a taxane. The cohort consisted of 85 patients from University Hospitals Cleveland Medical Center and 157 patients available publicly through the Cancer Imaging Archive.[27–29] Following chemotherapy, 48 patients achieved pCR and 194 retained the presence of residual disease (non-pCR). This cohort included patients from the ISPY1 (n=109) and Breast-NAC Pilot (n=48) studies that also had recurrence-free survival (RFS) information available. We considered RFS from the initiation of neoadjuvant chemotherapy (RFS for the Breast-NAC MRI Pilot study was recorded following completion of chemotherapy, but was adjusted based on the duration of treatment according to the study protocol[30]).

- *BRCA-TCHP:* A multi-institutional cohort of 129 HER2+ patients who received targeted neoadjuvant therapy at University Hospitals Cleveland Medical Center (n=28) or Cleveland Clinic (n=101) was also assessed. The majority of patients received neoadjuvant chemotherapy supplemented with trastumuzab and pertuzumab (n=125), while five patients from University Hospitals Cleveland Medical Center received only trastuzumab. 67 patients achieved pCR and 62 non-pCR. No BRCA-TCHP patients had survival information available.

**Lung:** A total of 187 standard dose, non-contrast lung CT volumes collected prior to treatment were included for analysis. Patients were treated and imaged at University Hospitals Cleveland Medical Center, and were divided into two groups depending on the type of therapeutic regimen that they received (i.e. trimodality or pemetrexed chemotherapy).

- *NSCLC-PLAT:* A total of 97 patients who received platinum-based chemotherapy without surgical intervention at Cleveland Clinic with available pre-treatment CT scans were identified. In the absence of post-treatment surgical samples, response was determined from imaging by RECIST criteria based on size changes between pre- and post-treatment CT. 47 patients were identified as responders, indicated by response or stable disease following platinum-based chemotherapy, while 49 had progression on imaging and were deemed non-responders. 92 patients had progression-free survival (PFS) information available, which was defined as the time from initiation of treatment to the detection of progressive disease or death, whichever occurred earlier, and was censored at the date of last follow-up for those alive without progression.
- *NSCLC-TRI:* 90 patients received trimodality therapy, consisting of neoadjuvant chemoradiation followed by surgical intervention. The endpoint for response was major pathologic response (MPR), defined as 10% or less residual viable tumor after neoadjuvant chemoradiation and the recommended surrogate endpoint in resectable NSCLC.[31] 36 patients achieved MPR. Recurrence free survival (RFS) was measured from the date of surgery to the date of recurrence or the date of death, whichever occurred earlier, and censored at the date of last follow-up for those alive without disease recurrence.

**Stratification.** For each treatment group, patients were divided into training and testing sets. Models were developed and optimized on the training set, then applied to the testing set. Three of the treatment groups (BRCA-TCHP, NSCLC-PLAT, NSCLC-TRI) had response rates of approximately 50%, and were accordingly divided randomly in half for training and testing, when possible using the same splits from prior studies.[32,33] Relative to these treatment strategies, the rate of response to BRCA-ACT is substantially lower.[34] Given the potential of training data imbalanced between categories to negatively impact classifier performance and robustness,[35] a BRCA-ACT training cohort was randomly chosen containing 50% of responders and enough non-responders to enforce a 3:1 class balance (previously shown to limit the negative effects of class imbalance for an LDA classifier[36,37]). The composition of each training and testing set, along with availability of response and survival endpoints, is summarized in Supplementary Table 1.

### *Quantifying the tumor-associated vasculature*

**Pre-Processing and Segmentation – Lung CT.** All lung CT volumes were resized to an isotropic resolution of 1 mm$^3$. Tumor boundaries were manually annotated in 3D by an experienced reader. Automatic segmentation of the tumor vasculature was then performed, as depicted in Supplementary

Figure 1. Next, the tumor-associated vasculature was extracted in several steps with a protocol previously shown to effectively segment pulmonary vasculature on non-contrast CT.[38,39] Each CT image was masked to the lungs by thresholding at a value of -550 HU followed by morphological processing[40] (Supplementary Figure 1b). Next, an open-source,[41] multi-scale 3D vessel enhancement filter[42] was applied to emphasize tubular vessel-like structures (see supplementary methods - implementation details for parameters), as illustrated in Supplementary Figure 1c. Thresholding was applied to the vessel enhancement image via Otsu's method[43] to isolate pixels belonging to the vasculature, then morphological operations were applied to remove noise and non-vessel artifacts (Supplementary Figure 1d). A box containing the tumor and an additional 5 cm in each direction was extracted for further analysis (Supplementary Figure 1e). An open-source fast marching algorithm[44] was applied to the segmented vasculature to identify the center lines of vessels[45] and divide the vessel network into discrete constituent branches (Figure 1b).

**Pre-Processing and Segmentation – Breast MRI.** The first post-contrast scan was spatially aligned to the pre-contrast scan via affine registration[46] and the difference in image intensities before and after contrast enhancement was then computed, yielding a subtraction image (Supplementary Figure 2a). Volumes were resized to an isotropic resolution of 1 mm$^3$. 3D tumor boundaries were obtained with a combination of manual annotation and automated segmentation techniques. First, partial tumor annotations on several adjacent axial slices were manually delineated by an experienced reader or derived from segmentations provided for publicly available data.[28,29] A 3-dimensional active contour segmentation algorithm[47] (the 'chenvese' function in MATLAB[47]) was applied to expand the annotated 2D slices to a full volumetric segmentation of the tumor in 3D. Vessel segmentation is depicted in Supplementary Figure 2. The heart and posterior torso were automatically detected and removed (Supplementary Figure 2b), and a vessel enhancement filter[41,42] was again applied (Supplementary Figure 2c) to detect vessel-like objects within the volume (see supplementary implementation details for parameters). Given the lack of true quantitative values in MRI as compared to CT, the vessel enhancement volume was segmented at multiple thresholds derived by Otsu's method,[43] which were each refined by morphological operations (Supplementary Figure 2d). The resulting segmentations (Supplementary Figure 2e) were assessed for alignment with vessel enhancement on maximum intensity projections and in 3D. The threshold that best captured the enhancing vasculature within each scan was selected manually by a single reader blinded to clinical data and therapeutic outcome for further analysis. Volumes were cropped 5 cm from the tumor in each dimension (Supplementary Figure 2f), and center line coordinates and branches of the final vessel network were computed by fast marching (Figure 1b).[23,44]

**Measures of QuanTAV Morphology:** From 3D vessel skeletons, 61 quantitative vessel tortuosity features, expanded from a set of 35 introduced previously,[17] were computed. The full set of QuanTAV Morphology features is described in Supplementary Table 2. At each point within a branch, curvature was computed as the inverse of the radius of the circle containing that point and the two adjacent points within the branch. Distribution of curvature was summarized along the entire vasculature and each branch through first order statistics (mean, standard deviation, max, skewness, and kurtosis), and branch-level statistics were summarized at the patient level with the same statistics. For each branch, torsion was computed as one minus the ratio of the Euclidean distance between the first and last points of a branch to the branch's length and summarized at the patient level via first order statistics. The distributions of curvature and torsion across the full vasculature were further summarized via 10-bin histograms. Additional vessel metrics – including vessel volume, length, number of vessels entering the tumor, and percentage of vessels in the vessel network feeding the tumor – were also computed.

**Measures of QuanTAV Spatial Organization:** A set of 30 features describing the organization of the tumor-associated vasculature, previously introduced[24] and listed in Supplementary Table 3, were computed. The steps for extracting QuanTAV Spatial Organization features are depicted in Supplementary Figure 3. From vessel centerlines, a set of 2D vessel projection images are generated, across which statistics summarizing the local orientation of vessels are computed. Along a projection image, the five most prominent vessel orientations are computed within a local window of fixed size via the Hough transform, a mathematical operation for the detection of lines within an image. The window of analysis is moved incrementally along the image to obtain a distribution of vessel orientations across the entire vessel image. The overall distribution of vessel orientations is then summarized by five first order statistics (mean, median, standard deviation, skewness, and kurtosis), which constitute the set of QuanTAV Spatial Organization features.

This process is applied to six distinct projection images. A set of three Cartesian projections is obtained by flattening the vasculature along one of the three spatial dimensions, in the axial, sagittal, or coronal planes. In addition to analyzing the TAV in the original coordinate system, each point within the 3D vasculature is also converted to a spherical coordinate system in order to capture vessel position relative to the tumor. Rather than (*X, Y, Z*) position, spherical coordinates correspond to *elevation* from the tumor center, *rotation* about the tumor center, and *distance* from the tumor surface. As with Cartesian views, the tumor vasculature is projected along each of these dimensions to obtain three 2D projection images: elevation with respect to rotation, rotation with respect to distance, and elevation with respect to distance. QuanTAV organization features are computed with two tunable parameters: maximum vessel distance from the tumor to include and size of the sliding window used to compute vessel orientations. These parameters were optimized within each imaging modality/cancer domain, and the process and results are described in greater detail within the expanded implementation details located in the supplementary methods.

## *Signature Development and Evaluation*

**QuanTAV Predictive Response score**. The set of top features that best predicted therapeutic response for each use case was identified in two rounds of Wilcoxon feature selection in 3-fold cross-validation within the training set. The size of this feature set was determined per cohort based on performance within the training set in cross-validation (see supplementary implementation details). For each cohort, top vessel features were incorporated into a linear discriminant analysis (LDA) classifier and trained across the full training cohort to predict response in the testing set. The output of this classifier was a score between 0 and 1, in turn corresponding to the level of confidence that a patient would achieve a response following the conclusion of therapy.

**QuanTAV Prognostic risk score and Groups**. For each cohort with survival information available, a survival model was derived in the training set to generate QuanTAV risk scores using a strategy inspired by Bhargava et al.[48] All observation times were censored at a maximum of 10 years. Features that were highly correlated and likely redundant were pruned from the feature set, retaining the feature with the highest absolute coefficient value in a multivariable proportional hazards model. A Cox regression model was trained using the remaining vessel features via 10-fold elastic net regularization using the Glmnet for MATLAB package.[49] The coefficient values for the model were then applied to training and testing sets to derive patient risk scores. A risk score threshold to optimally stratify patients into high and low risk groups based on maximizing the hazard ratio was derived in the training set for each cohort. Further implementation and optimization details are described in the supplementary methods.

**Statistical Analysis.** The primary metric used to evaluate response prediction models were odds ratio (OR) and area under the receiver operating-characteristic curve (AUC). Significance level and 95% confidence intervals of the AUC were computed via permutation testing with Monte Carlo sampling[50,51] across 50,000 iterations, described in detail in the supplementary material of a previous manuscript.[52] The univariable and multivariable association of QuanTAV response score and clinical variables with response were assessed based on OR in a logistic regression model. Clinical variables with univariable significance in the training set were incorporated into a clinical feature only logistic regression model, as well as a logistic regression model combining clinical variables with QuanTAV response score. The univariable and multivariable association of QuanTAV response score and clinical variables with response were assessed based on OR in a logistic regression model containing all features.

For prognostic models, both the QuanTAV risk score and categorical QuanTAV risk groups were assessed in univariable and multivariable settings along with baseline clinical variables, as well as pathological and treatment response information available at the completion of chemotherapeutic regimen. Cox proportional hazards models and risk groups were derived from only baseline clinical variables for comparison against QuanTAV risk groups. The primary metrics used to evaluate association with survival were hazard ratio (HR) and concordance index (C-index). Univariable and multivariable testing for significant association with prognosis was assessed based upon the coefficients of a Cox model within the cohort of interest.

### Data Availability

A portion of the data used in this study is publicly available through the Cancer Imaging Archive (TCIA).[27] Access to datasets from the University Hospitals Cleveland Medical Center (used with permission for this study) should be requested directly from these institutions via their data access request forms. Subject to the institutional review boards' ethical approval, unidentified data would be made available as a test subset.

## Results

### Predicting response and recurrence for anthracycline-based neoadjuvant chemotherapy (BRCA-ACT) from pre-treatment breast MRI

For the majority of breast cancer patients who receive neoadjuvant treatment, a chemotherapy-only regimen followed by surgery is standard-of-care.[53] A multi-institutional cohort of 242 patients who received anthracycline-cyclophosphamide alone or followed by a taxane (BRCA-ACT) was assembled and divided into subsets for training ($D_{tr}^{1}$) and independent testing ($D_{te}^{1}$). 19.8% achieved pathologic response on surgical samples following chemotherapy). 157 BRCA-ACT recipients additionally had 10-year recurrence-free survival (RFS) information available. Clinical details are summarized in Table 1.

Via cross-validation, a set of features discriminative of pathologic response were selected and used to train a classification model to yield a QuanTAV response score (Supplementary Table 4) that maximized performance in $D_{tr}^{1}$ (Supplementary Table 5). An increase in average torsion across vessels was identified as the feature most strongly associated with failure to achieve complete response (Figure 2, a&b). Torsion is defined as the complement of the ratio of the Euclidean distance between a vessel's start and end points to its total length, and is elevated in vessels with internal looping or "U"-shaped vessels that terminate near their origin.[13,54] The presence of such patterns in the vessels surrounding non-responsive tumors (Figure 2b) could impede the delivery of systemic therapy to the tumor and subsequently contribute to poor therapeutic response.

When applied to $D_{te}^{1}$, QuanTAV response score identified pathologic response with AUC=0.65 (95% CI 0.54-0.76, p=0.009) and was independently significant in a multivariable comparison with clinico-pathologic variables (Supplementary Table 6). Of three available clinical parameters, only hormone receptor positivity had univariable significance in $D_{tr}^{1}$. A model combining this variable and QuanTAV response score yielded an AUC=0.78 (95% CI 0.63-0.87, p=2e-5) in $D_{te}^{1}$, an increase over hormone receptor status only performance (AUC=0.69, 95% CI 0.58-0.80, p=0.0316). ROC curves for all models in $D_{te}^{1}$ are depicted in Figure 3a.

A regularized Cox proportional hazards model of QuanTAV features (Supplementary Table 7) was trained to derive a QuanTAV risk score via cross-validation in $D_{tr}^1$ (n=63). A risk score threshold for optimally stratifying patients into low- and high-risk groups was also derived in $D_{tr}^1$. Performance of QuanTAV risk score and risk groups in $D_{tr}^1$ and $D_{te}^1$ are listed in Supplementary Table 8. In $D_{te}^1$ (n=94), the model was significantly prognostic as both a continuous score (p=.0022, HR=1.25, 95% CI 1.08-1.44, C-index=.66) and categorical low- and high-risk groups (p=.0096, HR=4.25, 95% CI 1.29-14.07, C-index=.62). Despite only utilizing measurements from pre-treatment imaging, QuanTAV risk groups (Figure 3e) achieved similar prognostic performance to pathologic response on surgical sample after chemotherapy (Figure 3f).

We assessed the multivariable significance of QuanTAV risk predictions when compared with baseline clinical variables (age, size, and hormone receptor positivity) and functional tumor volume (FTV),[55] the volume of tumor that is actively vascularized based on contrast agent kinetics, which had previously been assessed for association with survival by Hylton et al. in data comprising a portion of the BRCA-ACT cohort. The QuanTAV model remained independently prognostic as both a continuous risk score (HR=1.20, 95% CI 1.04-1.40, p=0.014) and categorical risk groups (HR=5.51, 1.41- 21.49, p=0.014), along with the majority of clinical variables and FTV (Supplementary Table 9). We also assessed the correlation of each individual feature of the QuanTAV model with FTV. While several features were found to be significantly associated with FTV (Supplementary Table 10), such as features characterizing the quantity of vessels feeding the tumor, the large majority (10 of 14) were independent (p>0.05). This finding is consistent with the mutual independence observed between risk score and FTV (Supplementary Table 9), and suggests that QuanTAV provides prognostic information beyond clinical measures of perfusion.

### *Predicting response to neoadjuvant chemotherapy with targeted therapy for HER2+ breast cancers (BRCA-TCHP) from pre-treatment MRI*

Breast cancers with overexpression of the HER2 surface protein are highly aggressive, but can often be effectively combated through a targeted therapeutic strategy supplementing chemotherapy with monoclonal antibodies targeting the HER2 receptor. A second QuanTAV model was trained to predict response to a neoadjuvant regimen combining chemo- plus targeted therapy among HER2-positive breast cancer patients. The cohort (Table 1) consisted of 129 patients who were HER2-positive and received treatment with docetaxel, carboplatin, trastuzumab, and/or pertuzumab (TCHP), denoted as the BRCA-TCHP treatment group and divided into training ($D_{tr}^2$) and testing sets ($D_{te}^2$). Rate of pathologic response was 51.9%.

A QuanTAV response score model (Supplementary Table 11) was trained within $D_{tr}^2$ (Supplementary Table 5) to predict pathologic response to BRCA-TCHP. As was observed in BRCA-ACT, poor response to BRCA-TCHP was associated with elevated vessel torsion, as well as increased skewness of vessel orientations within the XY plane. Within $D_{te}^2$ (Figure 3b), the vessel model significantly predicted pathologic response (AUC=0.63, 95% CI 0.47-0.76, p=0.042). As a covariate in logistic regression models (Supplementary Table 12), however, QuanTAV response score was not significant in $D_{te}^2$ alone (OR=0.17, 95% CI 0.01-2.38, p=0.188). Hormone receptor status was significant in all subsets, with a testing set AUC of 0.64 (95% CI 0.52-0.75, p=0.017). Combining QuanTAV response score with hormone receptor status increased testing AUC to 0.70 (95% CI 0.53-0.82, p=0.0036).

### *Predicting post-treatment and long-term progression of NSCLC following platinum-based chemotherapy (NSCLC-PLAT) from pre-treatment CT*

In advanced NSCLC, platinum-based chemotherapy is standard of care first line treatment for patients lacking actionable mutations. The NSCLC-PLAT cohort consisted of 97 NSCLC patients who received a pemetrexed-based platinum doublet regimen and CT imaging before and after treatment at a single institution. In the absence of surgical samples, response was assessed on post-treatment

CT based on change from baseline in longest lesion diameter according to RECIST criteria.[56] 48.0% had responsive or stable disease on post-treatment imaging, and were categorized as responders, while the remaining patients experienced progression. 53 patients were used for training ($D_{tr}^3$) and 44 for testing ($D_{te}^3$).

The NSCLC-PLAT response score derived in $D_{tr}^3$ (Supplementary Table 5) consisted entirely of QuanTAV spatial organization features (Supplementary Table 13). According to this signature, progression was distinguishable by QuanTAV spatial organization features that corresponded to heterogeneous distribution of vessel orientations (particularly in the region immediately surrounding the tumor). While many NSCLC tumors shared high vascular density regardless of therapeutic outcome (Figure 4, a-d), QuanTAV spatial organization features reveal crucial architectural differences between responders (Figure 4b) and progressors (Figure 4d) at the tumor-vasculature interface. Vessel positions were converted to a spherical coordinate system (Figure 4, e&f), which were used to derive projection images of vessel organization relative to the tumor (Figure 4, g&h). For instance, elevated standard deviation of vessel orientations on projection images reflecting rotation and elevation with respect to distance from the tumor were strongly associated with progression (Figure 4h). Conversely, vessels surrounding responsive tumors maintained a consistent orientation towards the tumor's surface (Figure 4g).

When applied to $D_{te}^3$ (Figure 3c), QuanTAV response scores significantly predicted response on post-treatment imaging with AUC=0.70 (95% CI 0.54 – 0.85, p=0.024). Only age (p=0.048) and QuanTAV response score (p=0.010) significantly differed between responders and non-responders in $D_{tr}^3$. However, age was not predictive in $D_{te}^3$ (p=0.232) and did not improve the performance of the QuanTAV response score (Figure 3c). In contrast, QuanTAV response score was the only variable found to be independently significant (p<.045) in a multivariable comparison with six clinical variables in $D_{te}^3$ (Supplementary Table 14).

A QuanTAV risk score model (Supplementary Table 15) and corresponding low/high risk groups were derived in $D_{tr}^3$ (Supplementary Table 8) to predict progression-free survival (PFS): the time from initiation of chemotherapy until progression on imaging, metastasis, or death. Within $D_{te}^3$ (n=39), risk group (p=0.034, HR=2.29, 95% CI 1.07-4.94, C-index=0.62), but not risk score (p=0.141, HR=1.12, 95% CI 0.96-1.31, C-index=0.61), was significantly associated with PFS. When assessed for independence in a multivariable cox proportional hazards model (Supplementary Table 16) with clinical variables, QuanTAV risk group was the only variable found to be significant (p=0.028). KM plots for QuanTAV risk group and post-treatment RECIST response are depicted in Figure 3g and Figure 3h.

### Predicting response and recurrence to trimodality therapy (NSCLC-TRI) from pre-treatment CT

For patients with stage III resectable NSCLC, survival can be significantly improved by supplementing platinum-based chemotherapy with radiotherapy and surgical intervention,[57] known as tri-modality therapy and denoted here as NSCLC-TRI. 90 patients received pre-treatment CT, followed by neoadjuvant chemoradiation and surgery (Table 2). 41.1% of trimodality recipients achieved pathologic response, and longitudinal outcome data was available for all patients. Patients were divided randomly into training ($D_{tr}^4$) and held-out testing cohorts ($D_{te}^4$).

A NSCLC-TRI QuanTAV response score (Supplementary Table 17) was derived within $D_{tr}^4$ (Supplementary Table 5). In $D_{te}^4$ (Figure 3d), QuanTAV response score distinguished pathologic response with AUC=0.71 (95% CI 0.51-0.84, p=0.0093). Out of eight clinical and treatment-related variables, only Histology (adenocarcinoma vs. squamous cell carcinoma/other) was individually significant (p=0.0075) in $D_{tr}^4$ and predicted pathologic response with AUC=0.73 (95% CI 0.59-0.86, p=0.0002) in $D_{te}^3$. The combination of QuanTAV response score and histology outperformed either alone (AUC=0.85, 95% CI 0.69-0.94, p=2E-5). QuanTAV response score remained significantly associated with pathologic response in a multivariable comparison with all available clinical variables in $D_{te}^4$ (OR=.0004, 95% CI 0.00–0.18, p=0.012), as did histology (Supplementary Table 18).

Next, we assessed the capability of QuanTAV measures to predict RFS from date of surgery in recipients of trimodality therapy. A QuanTAV risk model (Supplementary Table 19) and corresponding risk groups were derived in $D_{tr}^4$ to stratify patients by RFS (Supplementary Table 8). Increases in the standard deviation of curvature across the length of the vessel was associated with elevated risk of recurrence (Figure 2, c), whereas tumors achieving durable response possessed fewer local variations in curvature due to bends and twists. Similar to the risk score derived for NSCLC patients receiving chemotherapy alone, QuanTAV Spatial Organization features measuring standard deviation of vessel orientation relative to the tumor centroid were also associated with recurrence or metastasis following surgery.

When applied to $D_{te}^4$, QuanTAV risk score (p=0.039, HR=1.28, 95% CI 1.01-1.62, C-index=0.66) and categorical risk groups (p=0.036, HR=3.77, 95% CI 1.09-13.00, C-index=0.64) were significantly prognostic. Kaplan meier curves illustrate the ability of pre-treatment QuanTAV risk groups (Figure 3i) and post-treatment pathologic response (Figure 3j) to stratify patients by RFS.

We assessed QuanTAV risk score and groups for independent prognostic value in $D_{te}^4$ in a multivariable comparison (Supplementary Table 20) including baseline clinical variables (age, sex, histology, clinical stage, largest lesion diameter, ECOG performance status, chemotherapy regimen, radiotherapy induction dose, and surgical procedure type), as well as features of pathology (vascular invasion and lymphatic invasion). Of these, only the QuanTAV model (p=0.037) and induction dose (p=0.035) were significant in a comparison with continuous risk score. Categorical QuanTAV risk group was similarly significant in a multivariable setting (p=0.013), along with several variables including induction dose, surgical procedure, lesion diameter, and presence of vascular invasion.

### *Assessing the robustness and generalizability of QuanTAV radiomics*

We sought to understand the impact of vessel segmentation errors on resulting QuanTAV models. To assess the robustness of our approach, we evaluated the performance of QuanTAV response scores in two testing sets (NSCLC-TRI, n=45 and BRCA-ACT, n=144) following various levels of perturbation to vessel masks (see supplementary methods - additional experiments). The vessel segmentations for each patient in the testing set were degraded through multiple iterations of randomized morphologic operations at each branchpoint and endpoint in the vessel skeletons (Supplementary Figure 4). Constituent QuanTAV features and corresponding response scores were recomputed from vessel segmentations after 5, 10, 15, and 20 iterations of perturbation. When QuanTAV response score generated from perturbed vessel segmentations to the original response score values via Delong's test of paired ROC curves,[58] no significant difference in AUC was found at any perturbation level in either the NSCLC-TRI (p=0.12-0.65) or BRCA-ACT (p=0.11-0.30) cohorts.

Additionally, we sought to assess the generalizability of QuanTAV analysis across institutions. Of the cohorts utilized in this study, only BRCA-ACT had sufficient data from multiple institutions to assess external generalizability. We conducted a post-hoc experiment after the finalization of our primary results within this cohort, where we repeated the training and validation of a pathologic response prediction model, but instead split our dataset according to its source: public (ISPY1-TRIAL and UCSF PILOT,[28,29] n=158) or private institution (University Hospitals, n=84). When QuanTAV response score was trained on public data and tested on private data, response prediction improved to AUC=0.71 (95% CI: 0.56-0.84, p=0.01). Similarly, when trained on private data and tested on public data, performance was slightly reduced to AUC=0.63 (95% CI: 0.51-0.72, p=0.006). These findings are consistent with our primary findings without separation by institution (AUC=0.65 [95% CI 0.54-0.76], p=0.009, n=144). Accordingly, we believe that QuanTAV model performs robustly across institutions, with only slight variations in performance that seem to correspond roughly with the number of patients used in training (training n=85, testing AUC=0.63; training n=98, testing AUC=0.65; training n=144, testing AUC=0.71).

# Discussion

In this study, we presented a novel radiomic biomarker that was associated with prognosis and treatment response for two different cancer and three different therapy types. This new category of computational imaging biomarkers leverages morphologic measurements of the twistedness and architecture of the tumor-associated vasculature. These vessel-based measurements were found to predict response and survival following intervention across two cancers, two imaging modalities, four chemotherapy-related treatment strategies, and a total of 558 patients. The construction of the tumor's vascular network through neo-angiogenesis plays a crucial role in the determination of patient outcomes by fostering a tumor microenvironment that promotes tumor progression and therapeutic resistance.[14,59] The structural abnormality of the resultant vasculature directly opposes successful therapeutic intervention, possibly owing to poorer delivery of therapeutic agents to the tumor bed,[19] while also encouraging the formation of hypoxic regions[60] that reduce efficacy of therapeutic agents and accelerate the development of drug resistant subclones.[61] Consistent with the known deleterious role of abnormal tumor vascularization,[10,11,59] we found that the expression of features reflecting erratic vascular shape and arrangement were predictive of poor response and elevated risk following chemotherapeutic intervention. Our findings suggest the critical role played by the tumor-associated vessel network across cancer domains in promoting therapeutic response and outcome.

In breast cancer, QuanTAV measurements on pre-treatment dynamic contrast-enhanced MRI predicted patient outcomes following neoadjuvant treatment with two standard-of-care therapeutic strategies in need of validated predictive markers.[7,53] QuanTAV response scores were developed to predict pathologic response following a chemotherapy-only regimen of BRCA-ACT and a BRCA-TCHP regimen of chemotherapy and targeted therapy for patients with a targetable HER2 receptor status. In recipients of BRCA-ACT, QuanTAV-derived models were shown to strongly predict response and recurrence-free survival independent of clinical variables including hormone receptor status: one of the few predictive markers available across this large and heterogeneous patient population. Additional biomarkers of response and survival for BRCA-ACT is of high clinical interest, since only roughly a quarter of recipients will achieve a complete pathologic response.[34] Our findings are consistent with prior research demonstrating that the formation of hook-like vessels feeding the tumor, known as adjacent vessel sign, is associated unfavorable prognosis and tumor phenotype.[62] We also investigated the ability of QuanTAV to predict response to a targeted BRCA-TCHP regimen. We observed that the BRCA-TCHP QuanTAV response score achieved a statistically significant ROC AUC and improved performance when incorporated into a clinical model, however it was not found to be independent as a logistic regression coefficient. In contrast to the other therapies explored in this work, HER2-targeted therapy is mechanistically anti-angiogenic[63,64] and helps normalize the TAV, thus potentially reducing the prognostic value of vascular shape and architecture within this treatment group.

These findings were mirrored in advanced NSCLC, where QuanTAV measures extracted from pre-treatment CT volumes were associated with both response and survival following two intervention strategies. First, for advanced NSCLC patients without actionable mutation, a platinum-based chemotherapy regimen is standard first-line intervention (NSCLC-PLAT). However, only 24%-31% of patients will achieve response and there are no clinically validated biomarkers for the guidance of platinum-based chemotherapy by benefit.[65] QuanTAV measures were predictive of response on post-treatment imaging according to RECIST criteria,[56] as well as progression-free survival. Second, for patients with stage III resectable NSCLC, survival can be significantly improved by a trimodality regimen supplementing platinum-based chemotherapy with radiotherapy and surgical intervention. However, trimodality therapy lacks predictive pre-treatment markers of benefit and bears a high rate of mortality between 5% and 15%.[57] Elevated disorganization and twistedness of the TAV on imaging was associated with a failure to achieve pathologic response and poorer 10-year recurrence-free survival. Our findings are in agreement with the crucial role of the TAV in NSCLC outcomes, evidenced by the importance of lymphovascular invasion[66] as a prognostic marker and the benefits of

TAV-normalization via anti-angiogenic therapy for many NSCLC patients.[67] The discriminability of QuanTAV in a regimen including radiotherapy is also consistent with the known role of abnormal vessel geometry in creating a low blood flow, poorly oxygenated tumor microenvironment that that facilitates radio-resistance.[68]

Critically, we observed that our measurements offered prognostic value independent of measures of functional volume on DCE-MRI (Supplementary Table 9 and Supplementary Table 10), suggesting that discriminative attributes of tumor vascular network architecture may not be captured by contrast agent-based perfusion imaging. Morphologic aberrations of the TAV on radiology have previously been shown to be elevated in the case of breast[16] and NSCLC[17] malignancy, as compared to benign lesions. Reduction in the tortuosity of the TAV on high-resolution brain MR angiography throughout treatment has been shown to be associated with favorable treatment outcomes in metastatic breast cancer.[69,70] Conversely, vessel tortuosity that has not normalized following treatment provide an earlier indication of treatment failure than monitoring tumor growth.[71] To our knowledge, this is the first study to date demonstrating the potential of 3D vascular morphology for predicting therapeutic outcomes prior to treatment, as well as the most comprehensive investigation of its role as potential predictive and prognostic biomarkers across cancers, imaging modalities and treatment types.

QuanTAV analysis represents a new addition to an expanding body of work suggesting the potential of quantitative imaging features mined from radiology to provide predictive biomarkers:[72] an approach known as radiomics. One of the most frequently deployed families of radiomic features is image texture, which quantifies the heterogeneity or spatial arrangement of image signals. Across numerous cancers and imaging modalities, texture-based features of the tumor and its environment have allowed for stratification of tumors into clinically significant biology- and outcome-associated groups.[73–75] In breast cancer, textural patterns of the tumor,[76,77] peri-tumoral surroundings,[52,78] bulk parenchyma,[79,80] and lymph nodes[81] on imaging has shown associations with risk and responsiveness to neoadjuvant therapy. Likewise in NSCLC, textural analysis of the tumor and peri-tumoral lung parenchyma has shown promise in predicting benefit of a number of therapeutic approaches, including chemoradiation with and without surgery,[32,33,82] targeted therapy,[83] and immunotherapy.[84,85] Consistently across both cancers, evidence suggests that elevated textural heterogeneity portends poor prognosis and increased risk of non-response.[52,73,78] Tortuous tumor vasculature plays an established role in fostering a heterogeneous, treatment resistant tumor microenvironment, and, in turn, that heterogeneity fuels further chaotic tumor angiogenesis.[19] Thus, a disorganized TAV may be to some degree intertwined with the development of a texturally complex tumor microenvironment on imaging that forms the basis of such prognostic radiomic signatures. To investigate a potential explanatory relationship between the TAV and prognostic texture signatures, we performed a comparison (Supplementary Table 21) of QuanTAV features and risk score with a previously published intra- and peri-tumoral texture-based risk score[33] that was derived within the same NSCLC-TRI cohort. QuanTAV and texture-based risk scores were found to be significantly correlated ($r=0.23$, p=0.030). Of the five most prognostic individual QuanTAV features, reduced variability of curvature along vessels was inversely correlated with texture-derived risk score ($r=-0.41$, p=0.0001). This result warrants additional study of the role of angiogenesis as a potential basis for image texture-based biomarkers. Despite potential interactions of vascularity and image texture, we crucially also found that QuanTAV was independent and complementary to texture-based analysis. When these two signatures were combined, they better stratified patients by RFS (C-index=0.70) than risk scores from texture (C-index=0.61) or vessel (C-index=0.66) features alone: suggesting the potential of computational vessel features to complement and improve traditional radiomic analysis. We also repeated a multivariable comparison including QuanTAV and texture risk scores along with the clinical variables previously examined in Supplementary Table 20. We found QuanTAV (p=0.02) risk score but not texture (p=0.08) to be independently prognostic in this setting.

Our findings suggest that patients with convoluted vasculature at time of treatment are less likely to derive benefit from systemic therapeutic intervention. Resultantly, patients flagged as non-responders based upon analysis of the TAV may benefit from anti-angiogenic therapy. In NSCLC, bevacizumab, an anti-antiangiogenic targeted therapy, in combination with chemotherapy provides therapeutic benefit by blocking the VEGF receptor, down-regulating tumor angiogenesis,[86–88] and facilitating delivery of other systemic therapeutics.[20] However, bevacizumab is currently prescribed conservatively in NSCLC due to its toxicity and current lack of validated predictive markers of therapeutic benefit.[88] QuanTAV measurements could potentially identify NSCLC patients who would benefit from vascular normalization through the addition of anti-VEGF therapy to their therapeutic regimen. The role of bevacizumab in breast cancer remains controversial, having been previously approved and subsequently revoked for treatment of metastatic breast cancer by the FDA due to safety concerns.[89] However, its use in the neoadjuvant setting in combination with chemotherapy has been shown to improve rate of pathologic response[90,91] and overall survival[92] in breast cancer subsets with specific receptor status and genotype. These results illustrate the important role of patient selection in success of vascular normalization in breast cancer and raise the question of whether anti-angiogenic therapies could still be an effective therapeutic option for these patients given more effective tools for targeting their application. Future work should explore the potential association of QuanTAV phenotype and benefit of anti-angiogenic therapy in NSCLC and breast cancer.

Our study did have its limitations. First, our segmentation protocol was formulated to achieve a balance of accuracy and efficiency in order to enable analysis within such a large cohort. To assess the robustness of our approach, we evaluated the performance of QuanTAV-based response scores in a breast MRI and lung CT dataset following various levels of disruption to vessel masks and found QuanTAV signatures to be robust to noise in vessel segmentations (Supplementary Figure 4). Beyond this experiment, it is encouraging that even accounting for segmentation errors, our approach was found to be predictive and prognostic in a wide number of use cases. We did not explore more sophisticated methods of isolating the tumor vasculature in this work, such as specialized deep learning segmentation strategies.[93,94] However, we have shown in a subsequent, preliminary study[95] that fully automated deep learning-based vessel segmentation also enables prognostic QuanTAV analysis in an additional disease and treatment domain – liver metastases treated with CDK 4/6 inhibitors – and validated this signature across institutions. Future work should compare various strategies of vessel segmentation in the context of QuanTAV performance. Second, breast MRI datasets were assembled across institutions and trials, and, consequently, imaging data was highly heterogeneous in acquisition protocol – a confounder we attempted to minimize through preprocessing strategies. Encouragingly, unsupervised clusterings of top QuanTAV features did not reveal site-based batch effects in either breast cancer cohort (Supplementary Figure 5). Third, across the datasets analyzed, it was necessary to utilize different clinical endpoints for response (pathologic complete response [pCR] for BRCA-ACT and BRCA-TCHP, major pathological response [MPR] for NSCLC-TRI, and RECIST response for NSCLC-PLAT) and survival (RFS for BRCA-ACT and NSCLC-TRI and PFS for NSCLC-PLAT), due to differing accepted and feasible clinical endpoints in the various clinical contexts (See Methods for further definition). For instance, pathologic response could not be assessed in NSCLC-PLAT since patients did not receive surgery. Finally, further validation of our approach is required in a prospective setting prior to clinical adoption. QuanTAV-based measurements should next be evaluated for their ability to predict well-defined clinical endpoints such as pathologic response among patients enrolled in clinical trials including chemotherapy.

## Acknowledgements:

Research reported in this publication performed by Nathaniel Braman was supported by:

- the National Cancer Institute under award number F31CA221383-01A1
- the National Institute for Biomedical Imaging and Bioengineering under award numbers T32EB007509
- the Hartwell Foundation

Research reported in this publication performed by Prateek Prasanna was supported by the National Cancer Institute under award number 1R21CA258493-01A1.

Research reported in this publication performed by Anant Madabhushi was supported by:
- the National Cancer Institute under award numbers 1U24CA199374-01, R01CA202752-01A1, R01CA208236-01A1, R01CA216579-01A1, R01CA220581-01A1, 1U01CA239055-01, 1U01CA248226-01, 1U54CA254566-01
- the National Heart, Lung and Blood Institute under award number 1R01HL15127701A1
- the National Institute for Biomedical Imaging and Bioengineering under award numbers 1R43EB028736-01
- the National Center for Research Resources under award number 1 C06 RR12463-01
- the National Institute of Diabetes and Digestive and Kidney Diseases through the Kidney Precision Medicine Project (KPMP) Glue Grant
- the United States Department of Veterans Affairs Biomedical Laboratory Research and Development Service under VA Merit Review Award IBX004121A
- the Office of the Assistant Secretary of Defense for Health Affairs, through
  - the Breast Cancer Research Program (W81XWH-19-1-0668)
  - the Prostate Cancer Research Program (W81XWH-15-1-0558, W81XWH-20-1-0851)
  - the Lung Cancer Research Program (W81XWH-18-1-0440, W81XWH-20-1-0595)
  - the Peer Reviewed Cancer Research Program (W81XWH-18-1-0404)
- the Ohio Third Frontier Technology Validation Fund
- the Clinical and Translational Science Collaborative of Cleveland (UL1TR0002548) from the National Center for Advancing Translational Sciences (NCATS) component of the National Institutes of Health and NIH roadmap for Medical Research
- The Wallace H. Coulter Foundation Program in the Department of Biomedical Engineering at Case Western Reserve University

The content is solely the responsibility of the authors and does not necessarily repressent the official views of the National Institutes of Health, the U.S. Department of Veterans Affairs, the Department of Defense, or the United States Government.

# Tables

|  | **BRCA-ACT** | **BRCA-TCHP** |
|---|---|---|
| **N** | 242 | 129 |
| **Pathologic response** | | |
| *Responder* | 66 | 66 |
| *Non-responder* | 176 | 63 |
| **Recurrence-free Survival** | | |
| *Event observed, N [median time-to-event, months]* | 48 [25.0] | N/A |
| *Censored, N [median time-to-last-followup, months]* | 109 [53.3] | N/A |
| *Unavailable* | 85 | 129 |
| **HER2 status** | | |
| *Positive, N* | 0 | 129 |
| *Negative, N* | 242 | 0 |
| **Hormone Receptor status** | | |
| *Positive, N* | 152 | 86 |
| *Negative, N* | 90 | 43 |
| **Age, years** | 49.6 ± 10.5 | 50.1 ± 11.1 |
| **Longest diameter, mm** | 5.4 ± 2.9 | 4.4 ± 2.9 |
| **Cohort** | | |
| *Training* | 98 | 69 |
| *Testing* | 144 | 60 |

*Table 1.* Breast cancer cohorts explored in this study: anthracycline-based neoadjuvant chemotherapy (BRCA-ACT) for HER2-negative patients and targeted therapy for HER2+ breast cancers (BRCA-TCHP).

|  | NSCLC-PLAT | NSCLC-TRI |
|---|---|---|
| **N** | 97 | 90 |
| **Pathologic Response** | | |
| *Major pathologic response (MPR)* | N/A | 37 |
| *non-MPR* | N/A | 53 |
| **Imaging Response** | | |
| *Response/Stable Disease* | 50 | 36 |
| *Progressive Disease* | 47 | 54 |
| **Survival** | | |
| *Event type* | Progression | Recurrence |
| *Event observed, N [median time-to-event, months]* | 75 [1.0] | 37 [17.6] |
| *Censored, N [median time-to-last-followup, months]* | 17 [36.0] | 53 [40.0] |
| *Unavailable* | 5 | 0 |
| **Histology** | | |
| *Adenocarcinoma, N* | 70 | 65 |
| *Squamous Cell Carcinoma/Other, N* | 27 | 25 |
| **Stage** | | |
| *I, N* | 4 | 0 |
| *II, N* | 1 | 0 |
| *III, N* | 20 | 90 [84 IIIA, 6 IIIB] |
| *IV, N* | 69 | 0 |
| *N/A, N* | 3 | 0 |
| **Sex** | | |
| *Male, N* | 51 | 49 |
| *Female, N* | 43 | 41 |
| *N/A, N* | 3 | 0 |
| **Smoking history** | | |
| *Past smoker, N* | 79 | -- |
| *Non-smoker, M* | 15 | -- |
| *N/A, N* | 3 | 90 |
| **ECOG Performance Status** | | |
| *0, N* | -- | 17 |
| *1, N* | -- | 66 |
| *N/A, N* | 97 | 7 |
| **Age, years** | 61.0 ± 13.0 | 63.2 ± 10.6 |
| **Longest diameter, mm** | 46.3 ± 31.7 | 52.2 ± 30.2 |
| **Cohort** | | |
| *Training* | 53 | 44 |
| *Testing* | 44 | 46 |

***Table 2.*** *Non-small cell lung cancer (NSCLC) cohorts explored in this study: platinum-based chemotherapy (NSCLC-PLAT) and trimodality therapy (NSCLC-TRI).*

# Figure Legends

**Figure 1.** Overview of the development and validation of quantitative tumor-associated vasculature (QuanTAV) response and risk scores. Models were trained and validated in four therapeutic cohorts: anthracycline-based neoadjuvant chemotherapy [BRCA-ACT] and neoadjuvant chemotherapy with anti-HER2 agents [BRCA-TCHP] in breast cancer and platinum-based chemotherapy only [NSCLC-PLAT] and neoadjuvant chemoradiation with surgery [NSCLC-TRI] in non-small cell lung cancer. The tumor and associated-vasculature were extracted from pre-treatment breast DCE-MRI and chest CT. For each vessel network, centerlines were derived and two categories of QuanTAV features were computed: Morphology and Spatial Organization. QuanTAV morphology features quantified the shape of tumor vessels. Statistics describing the distribution of metrics such as curvature (inversely proportional to the radius of a circle fitting three adjacent vessel points) and torsion (detecting differences in vessel length relative to the distance between its start and end points) comprised the bulk of QuanTAV morphology features. QuanTAV Spatial Organization features evaluate the architecture of the vessel network by evaluating the degree of vessel alignment along 2D projection images depicting the position of vessels in either the imaging space (cartesian) or a coordinate system relative to the tumor center and surface (spherical). QuanTAV features were optimized to predict response in each training cohort, then a linear discriminant analysis (LDA) classifier was trained to predict response from a limited set of features selected by wilcoxon rank sum test. The classifier's was the QuanTAV response score, which was assessed for independent ability to predict therapeutic response in the testing set. Likewise, in the three cohorts with progression- (NSCLC-PLAT) or recurrence-free survival (BRCA-ACT, NSCLC-TRI) data available, a regularized cox proportional hazards model was trained to derive a QuanTAV risk score and low/high risk groups in the training set, which were assessed for univariable and multivariable association with survival in the testing set.

**Figure 2.** QuanTAV morphology measures detect differences in vessel shape on pre-treatment breast MRI and lung CT predictive of outcome following treatment including chemotherapy. Top: Elevated vessel torsion is associated with non-response to anthracycline-based chemotherapy in breast cancer (BRCA-ACT). a) maximum intensity projections of pre-treatment DCE-MRI subtraction images for patients who did (left) and did not (right) experience pathologic response following BRCA-ACT. b) Vessel torsion on pre-treatment dynamic MRI distinguishes non-responders and complete responders. For each discrete vascular branch, all corresponding voxels within the branch are shaded according to the torsion value of the branch. The vasculature of patients who do not respond (left) exhibit elevated torsion, indicating vessels that twist back on themselves and are more convoluted in shape. Conversely, patients who achieve pathologic response exhibit less tortuous vasculature that transports blood more directly towards the tumor or throughout the breast. Bottom, c) Curvature across vessels on pre-treatment CT differs between NSCLC patients who will (right) and will not (left) recur following neoadjuvant chemoradiation followed by surgery (NSCLC-TRI). Vessel center-lines are shaded according to local curvature, computed for every set of three adjacent points along a vessel. Elevated standard deviation of curvature was associated with recurrence following NSCLC-TRI, visible as regions of local bends and twists along the length of a vessel (right). Responsive patients, in contrast, were surrounded by vessels with fewer of these micro-deviations.

**Figure 3.** (a-d) Receiver operating characteristic (ROC) curves for the QuanTAV response score (blue), clinical model (red), and combined QuanTAV and clinical model (green) in testing sets for the four treatment cohorts. a) Prediction of pathological response for breast cancer patients receiving anthracycline-based neoadjuvant chemotherapy (BRCA-ACT, n=144). b) Prediction of pathological response for breast cancer patients receiving HER2-targeted neoadjuvant chemotherapy (BRCA-TCHP, n=69). c) Prediction of response on post-treatment imaging for NSCLC patients receiving platinum-based chemotherapy (NSCLC-PLAT, n=44). d) Prediction of pathologic response for NSCLC patients receiving a trimodality regimen of chemoradiation followed by surgery (NSCLC-TRI, n=44). (e-j) Kaplan Meier curves showing pre-treatment QuanTAV risk groups and post-treatment response. Note that QuanTAV risk groups are shown for the testing set, while response is shown for the full patient population used in this study. e,f) BRCA-ACT: association of QuanTAV risk groups (e) and post-treatment pathologic complete response (f) with 10-year recurrence-free survival. (g,h) NSCLC-PLAT: association of QuanTAV risk groups (g) and post-treatment RECIST response (h) with 10-year progression-free survival. (i,j) NSCLC-TRI: association of QuanTAV risk groups (i) and post treatment major pathologic complete response (j) with 10-year recurrence-free survival.

**Figure 4.** Organization of vascular network at the tumor interface distinguishes NSCLC tumors that experience durable response (left) from those that progress (right) following platinum-based chemotherapy (NSCLC-PLAT). High vascular density is observed in both non-progressors (a) and progressors (c), but differences in the arrangement of tumor-adjacent (b and d) vessels are detectable through QuanTAV Spatial Organization features. e-h) On projection images depicting rotation around the tumor centroid (left) and elevation above the tumor centroid (right) with respect to distance from the tumor, the standard deviation of vessel orientation was elevated among patients who experienced progression. The position of vessels in a spherical coordinate system relative to the tumor, depicted in polar plots for responders (e) and non-responders (f), were used to derive corresponding spherical projection map images (g and h, respectively). Vessel orientation is computed across projection images locally via a sliding window. Tumors that achieve durable response possessed orderly vasculature with linear paths towards the tumor (e & g). However, patients who experienced disease

progression possessed tumor-adjacent vasculature with twists and deflections from the tumor with respect to distance from its surface (f), quantifiable as increased standard deviation of orientation on spherical projection images (h). This abnormal vascular architecture may contribute to poor therapeutic outcome by constraining delivery of chemotherapeutics and promoting a treatment resistant tumor microenvironment.

Figure 1.

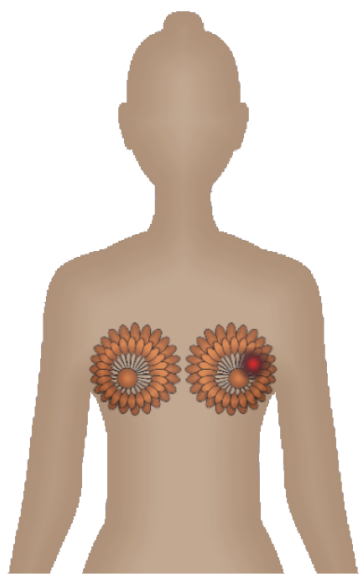
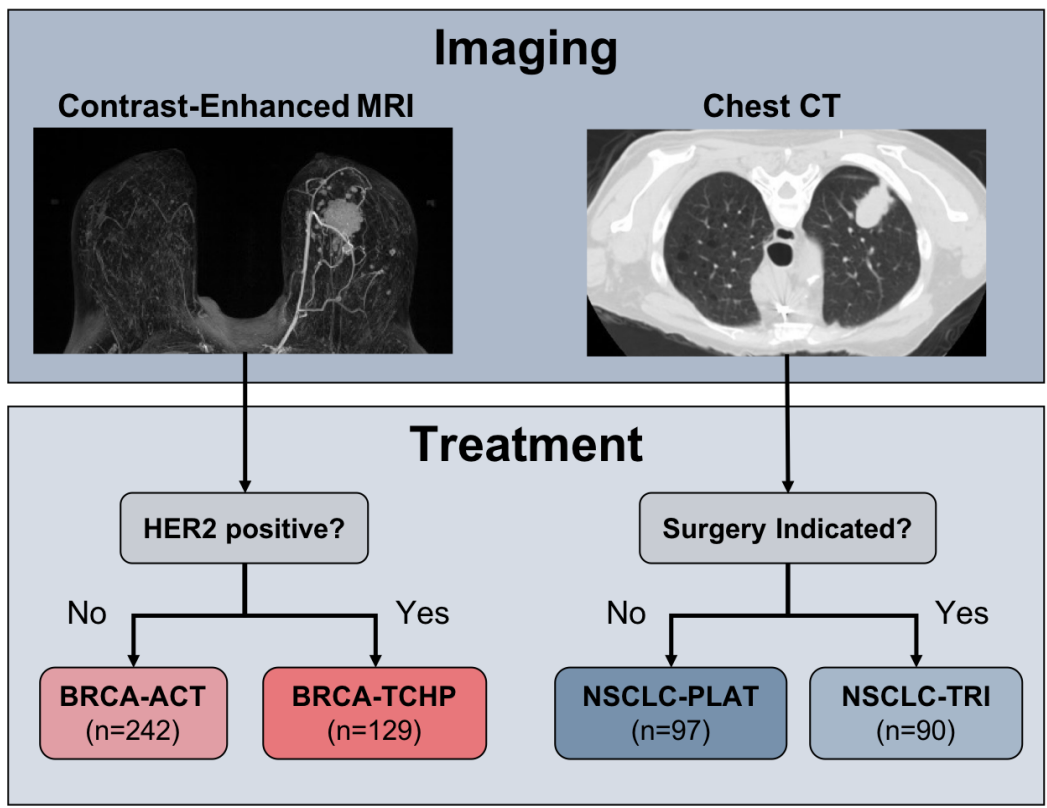
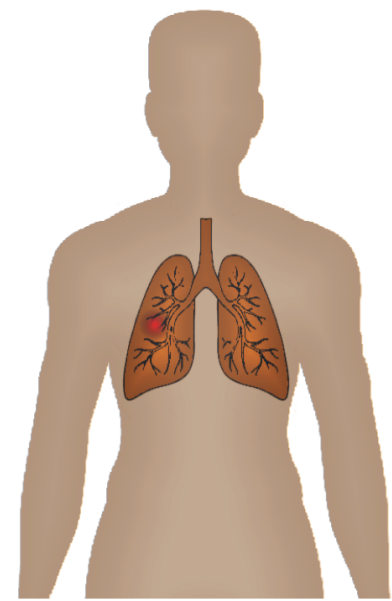
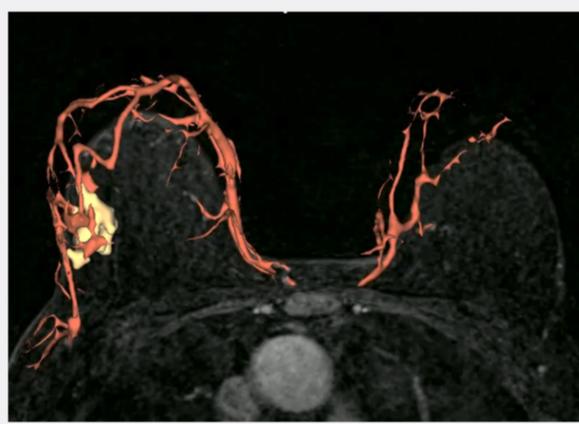
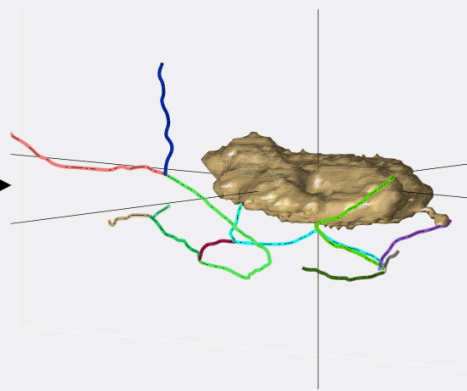
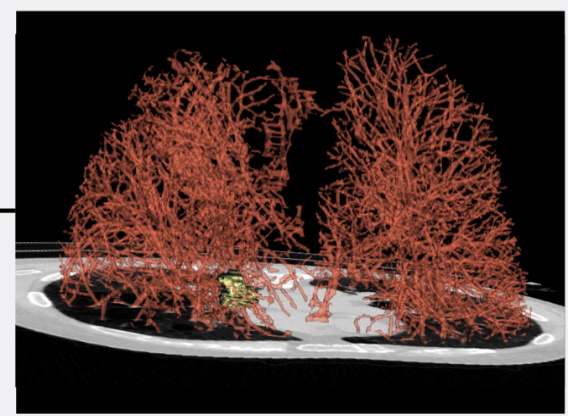
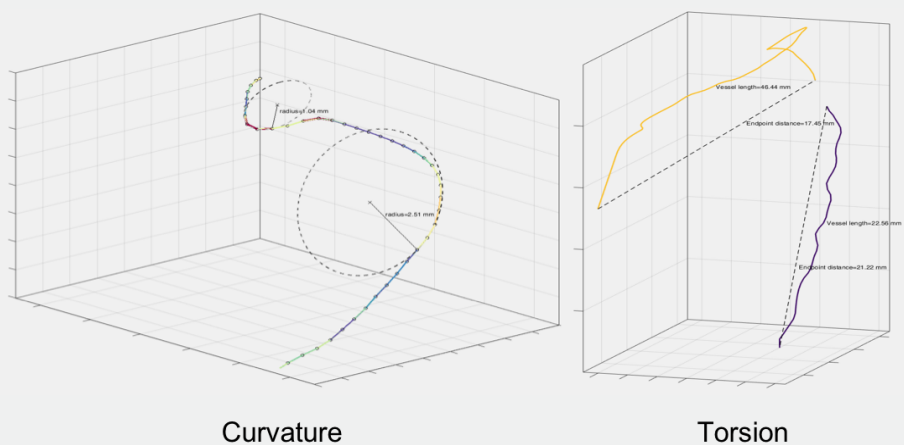
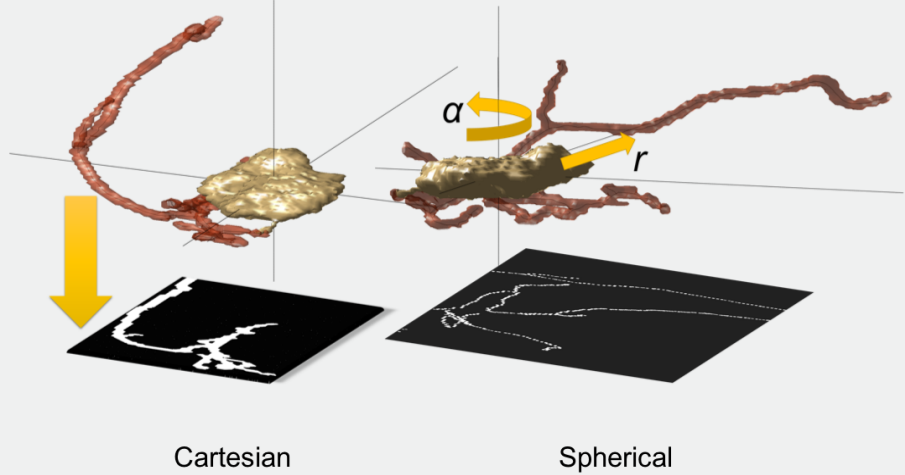
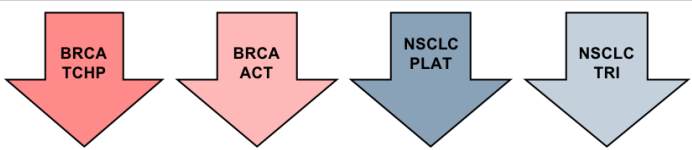
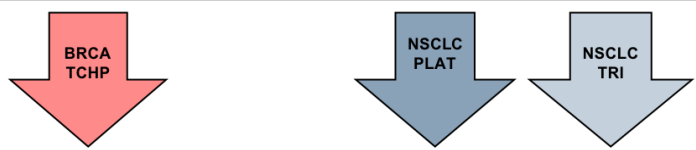
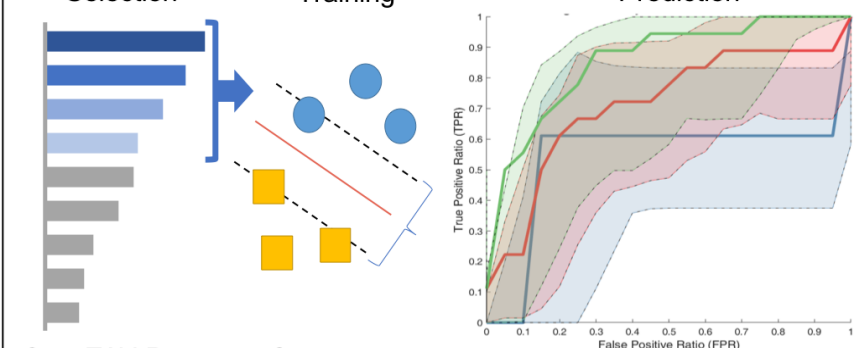
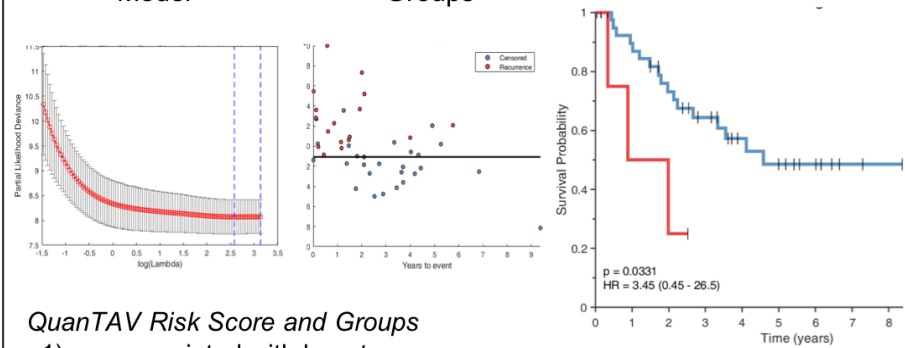

Figure 2.

**Pathologic Response**      **Non-Response**

**BRCA-ACT**: Vessel Torsion

a)
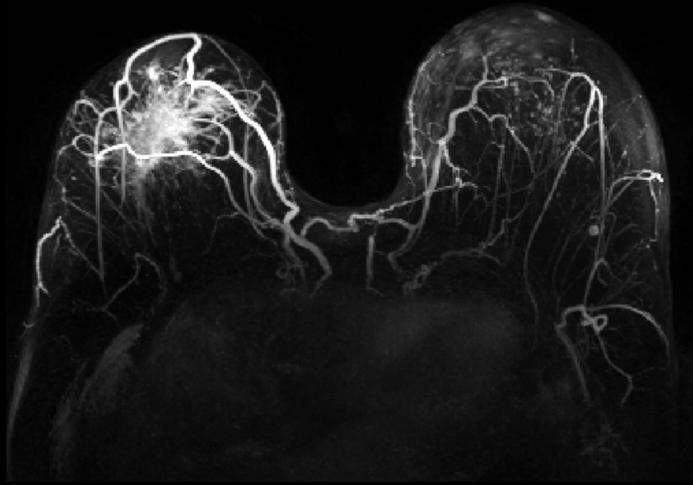
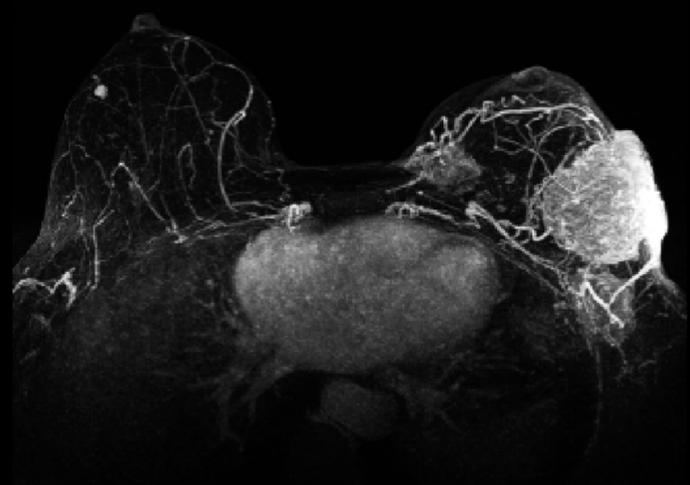

b)
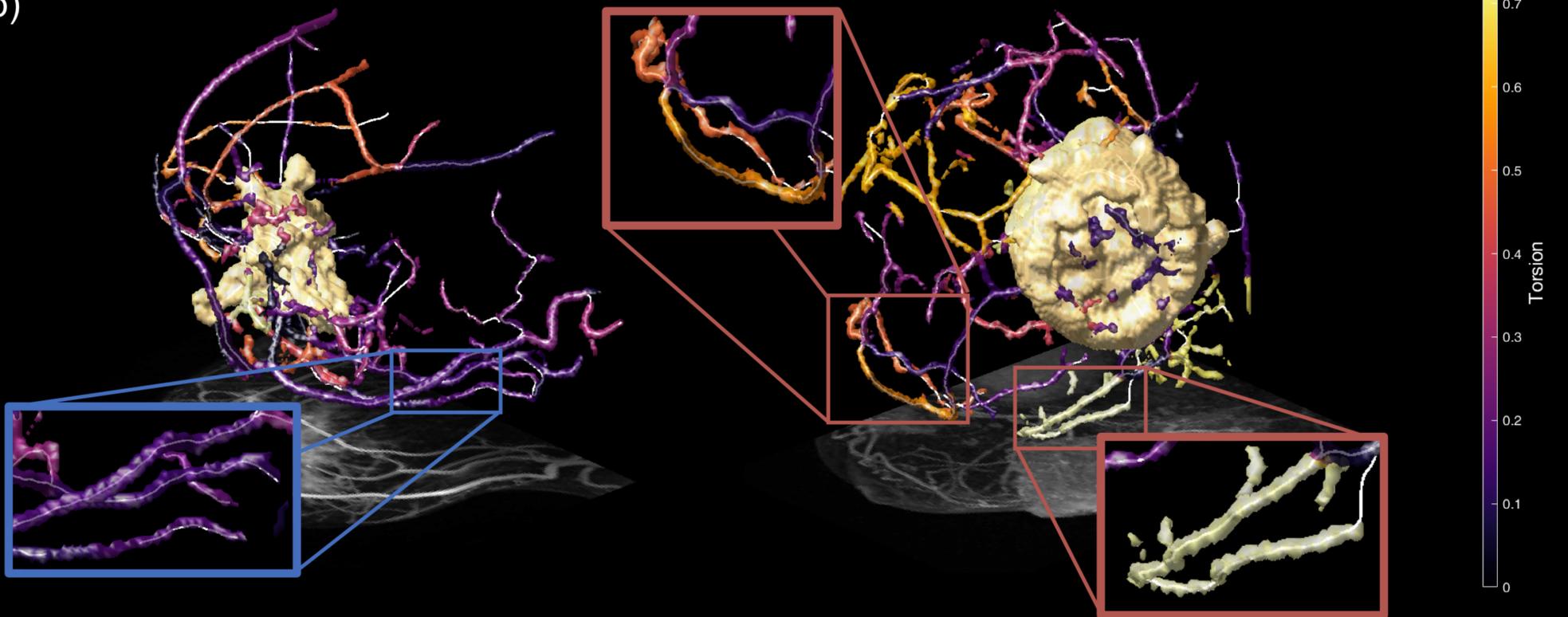

**NSCLC-TRI**: Vessel Curvature

c)    **Non-recurrence**      **Recurrence**
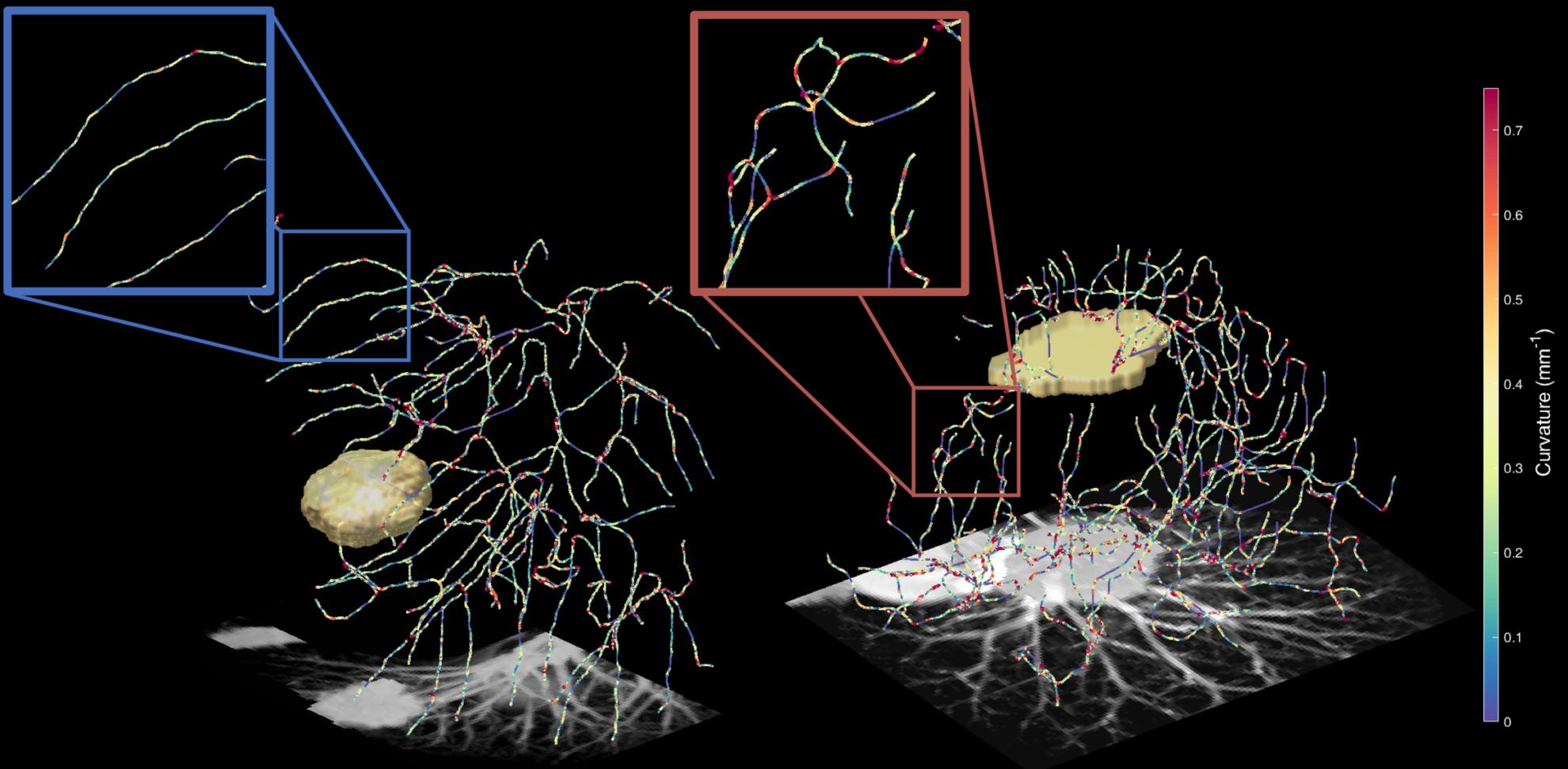

# Figure 3.

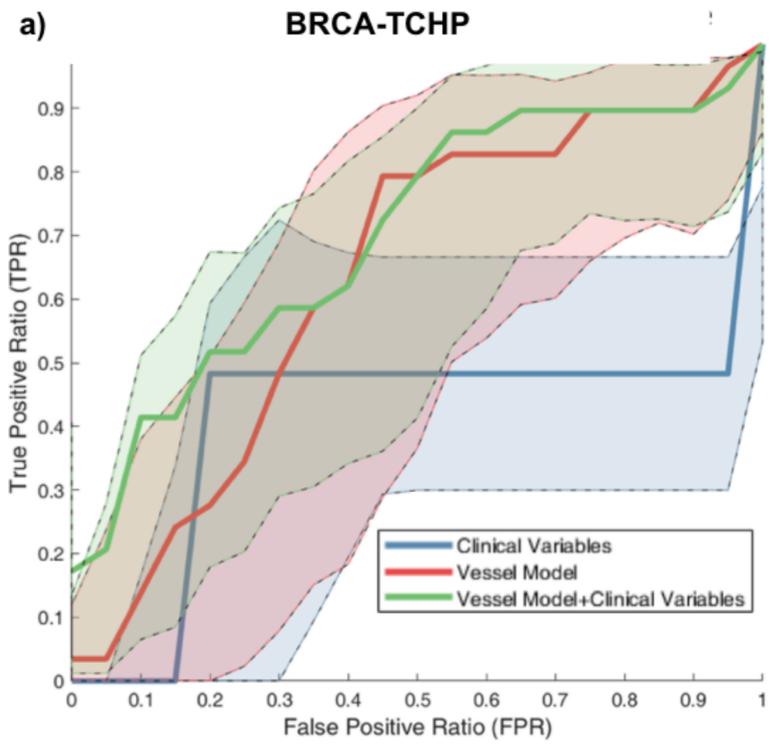 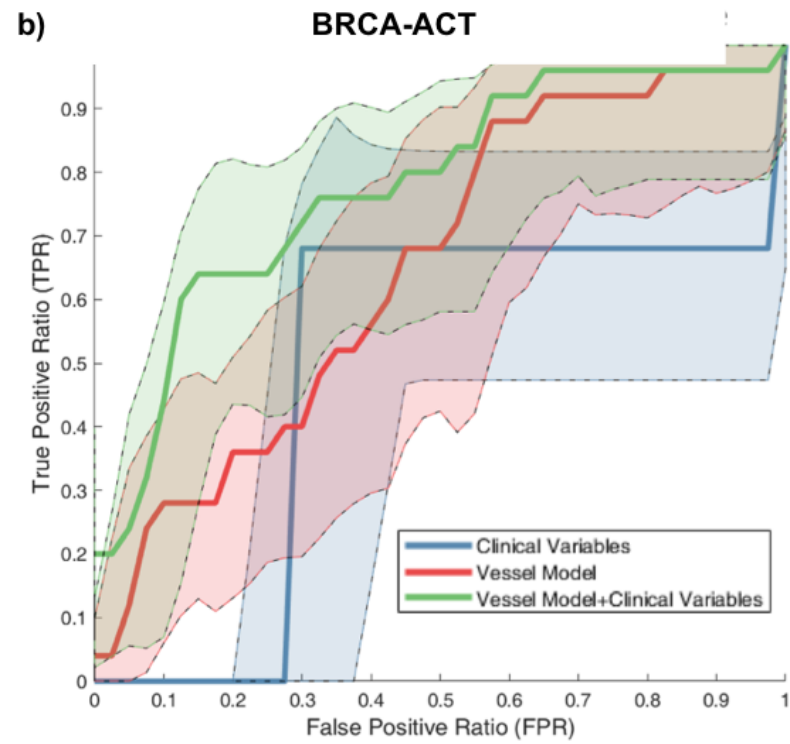 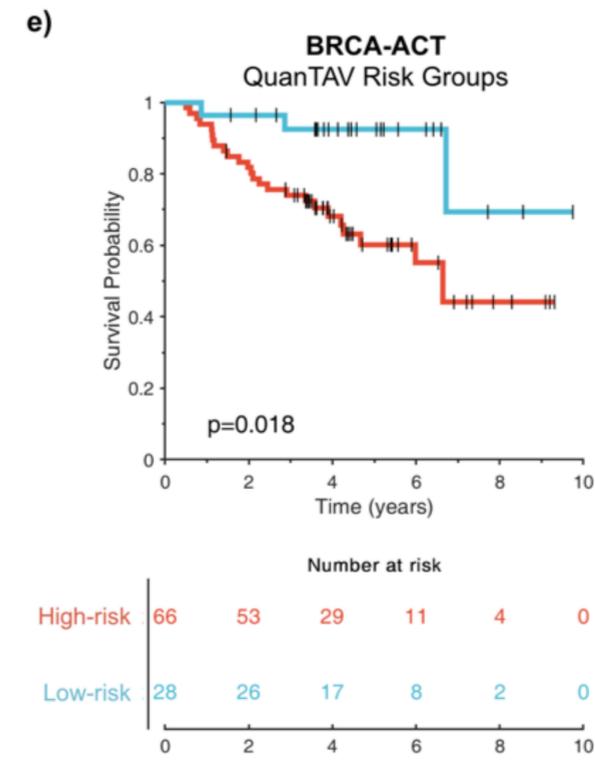 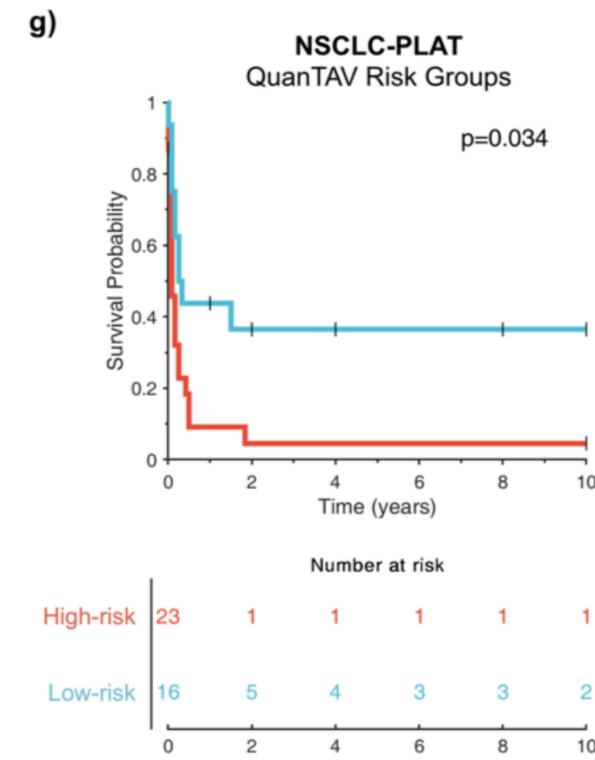 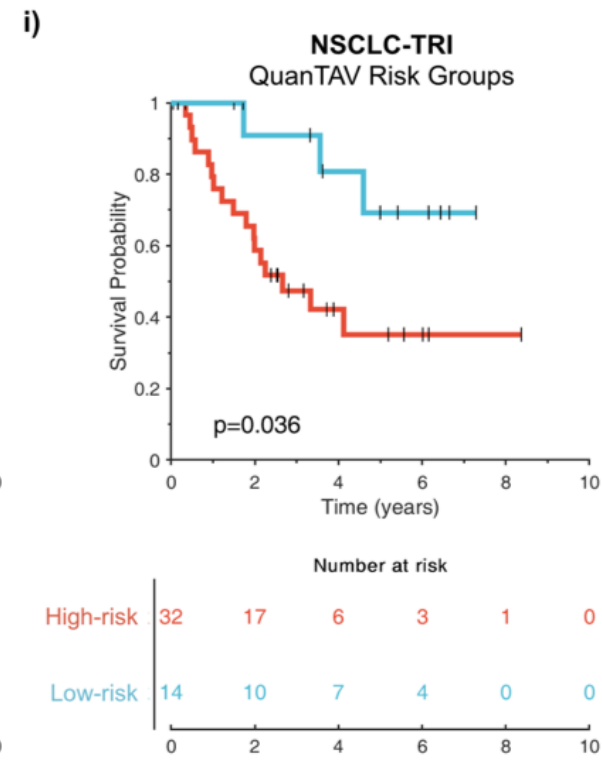

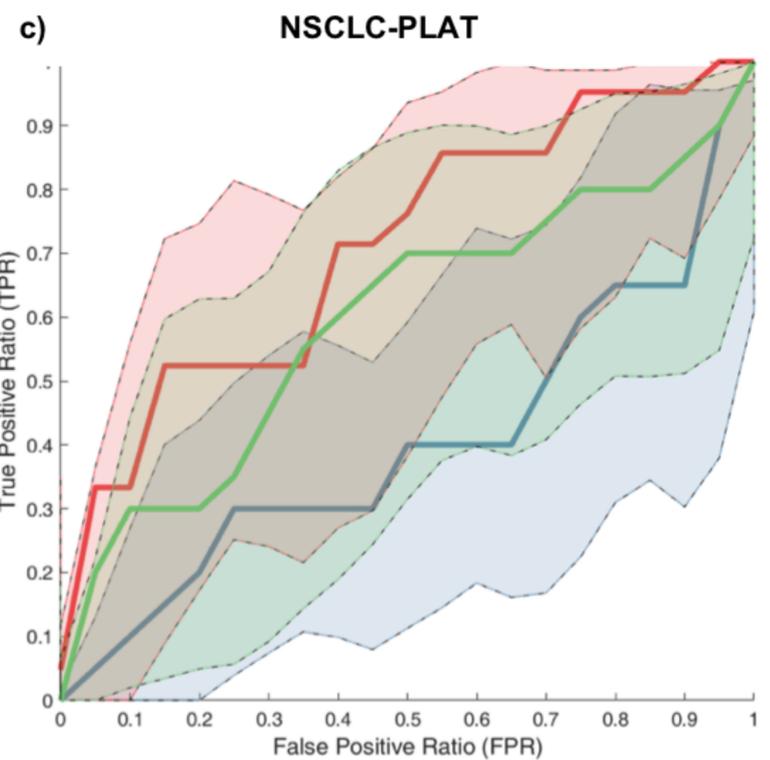 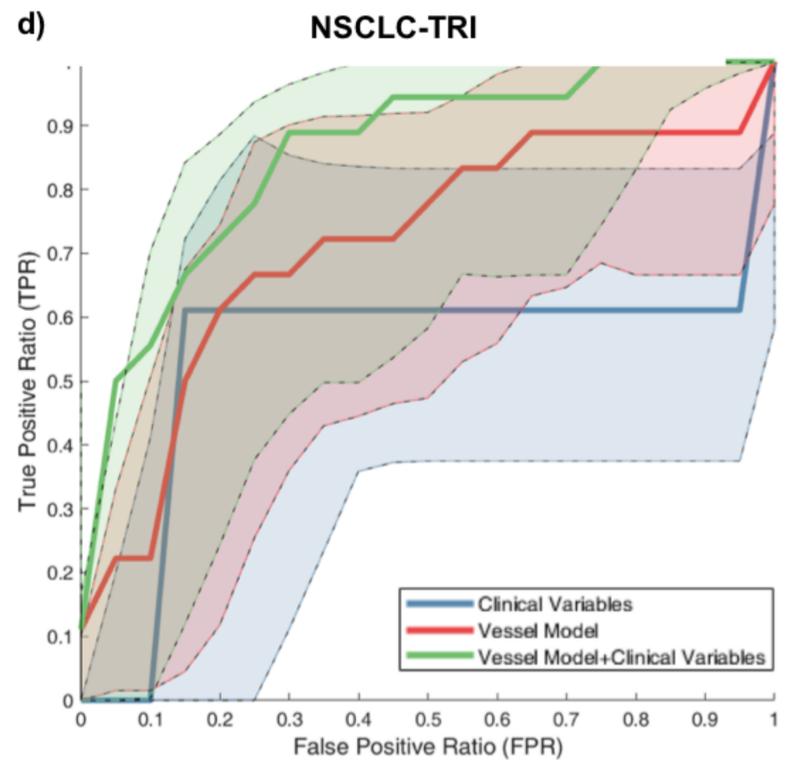 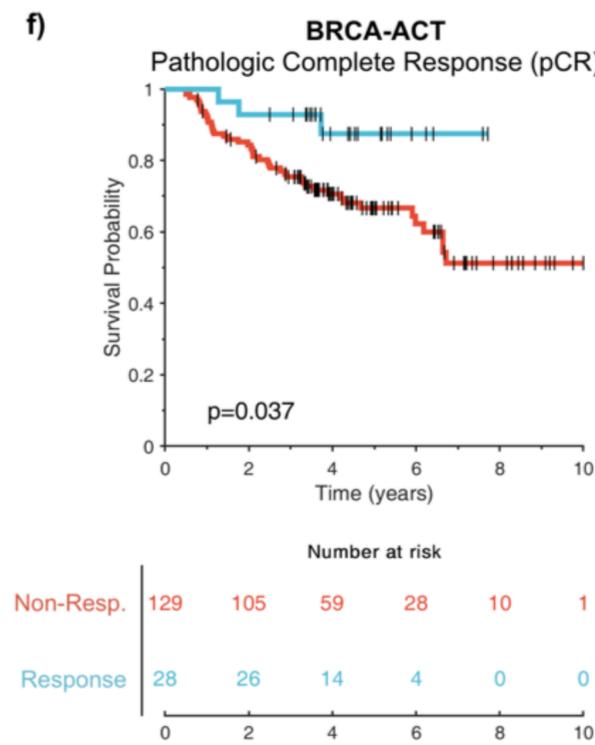 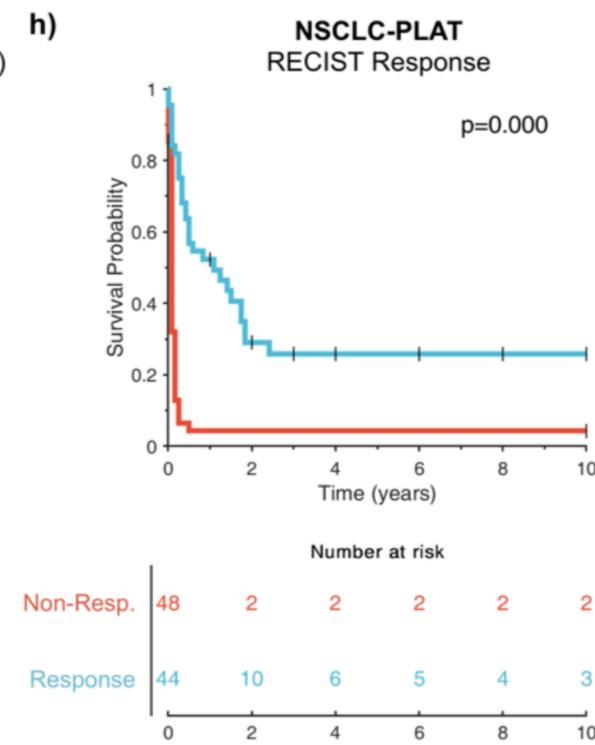 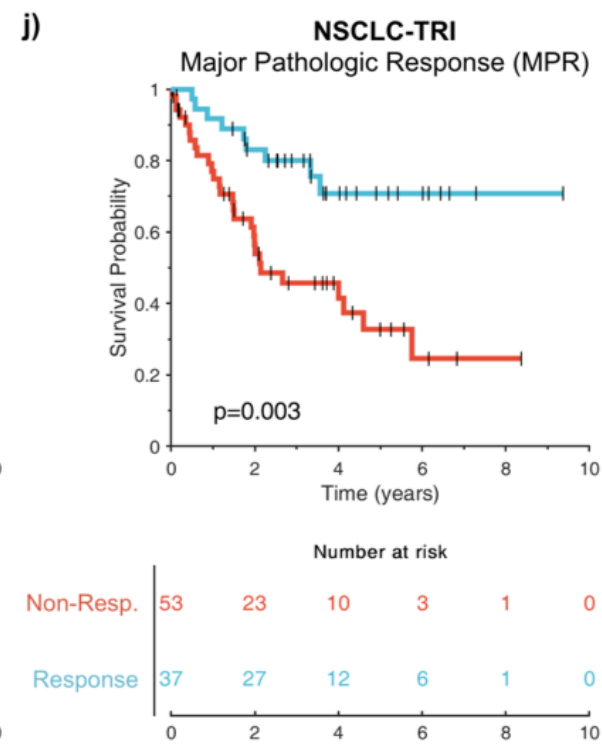

# Figure 4.

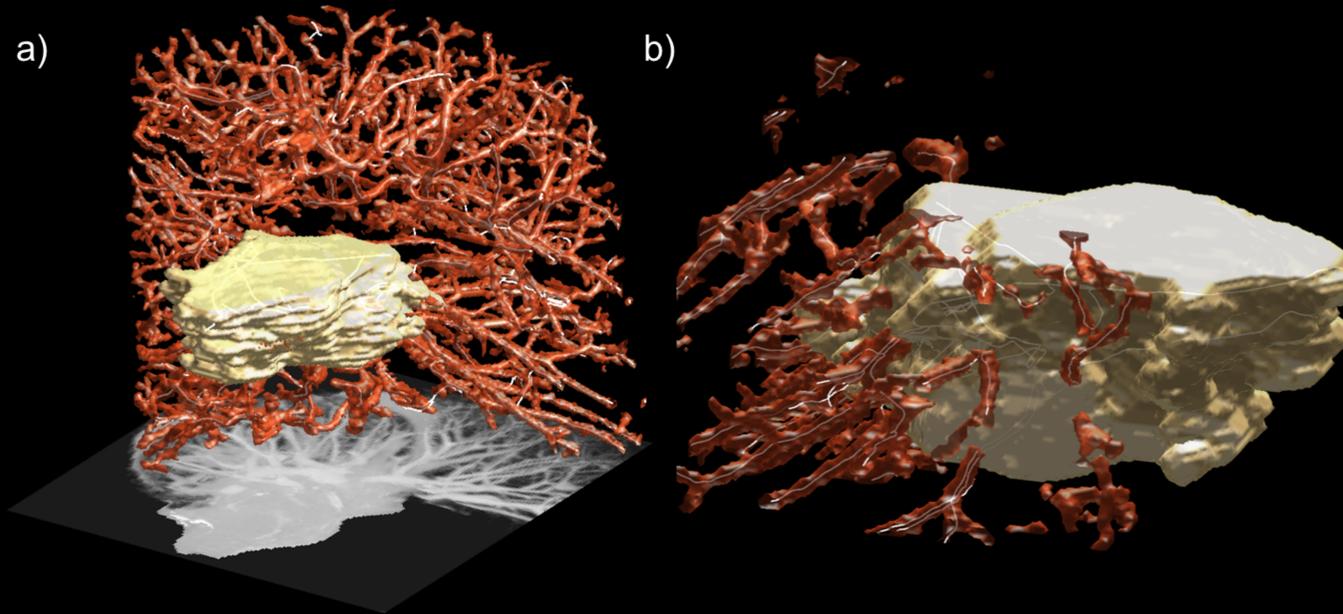
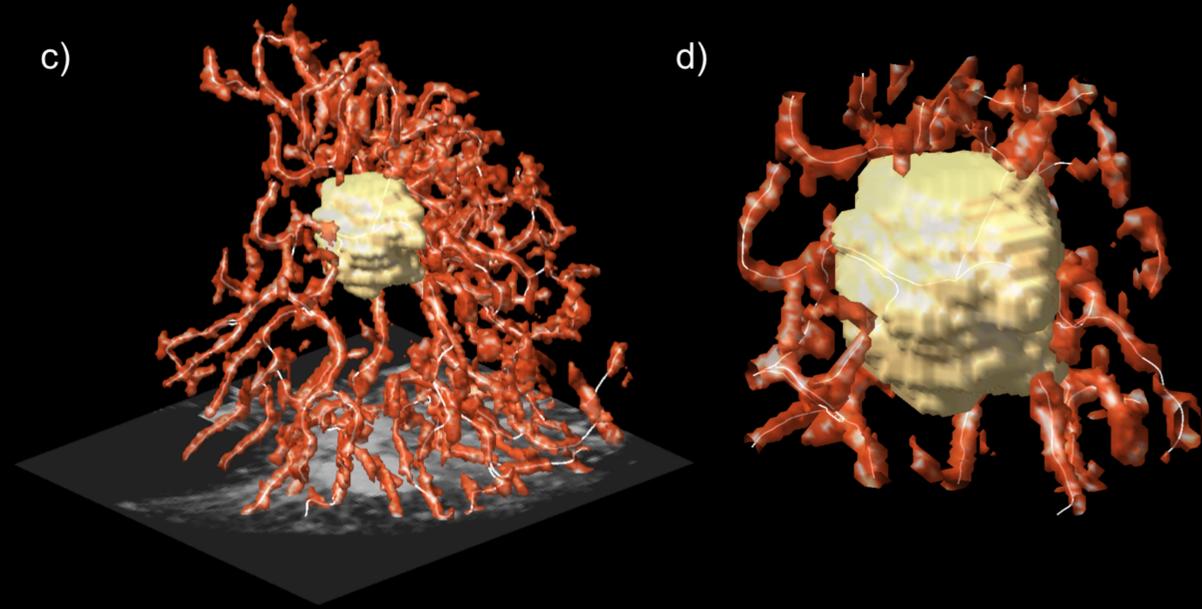

**Non-progression**     **Progression**

a) b) c) d)

## Spherical Coordinate Plots

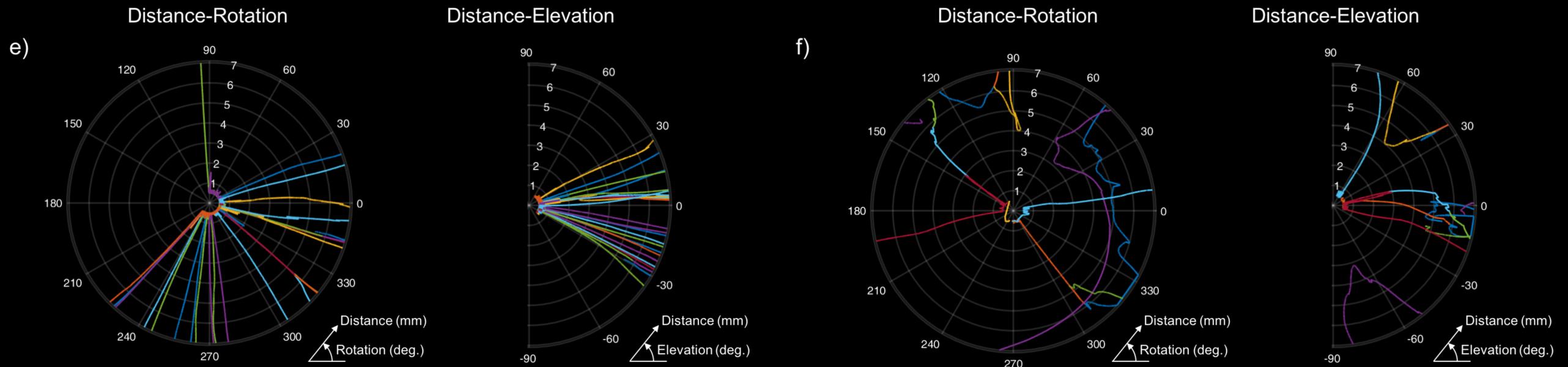

e) Distance-Rotation   Distance-Elevation    f) Distance-Rotation   Distance-Elevation

## Projection Maps

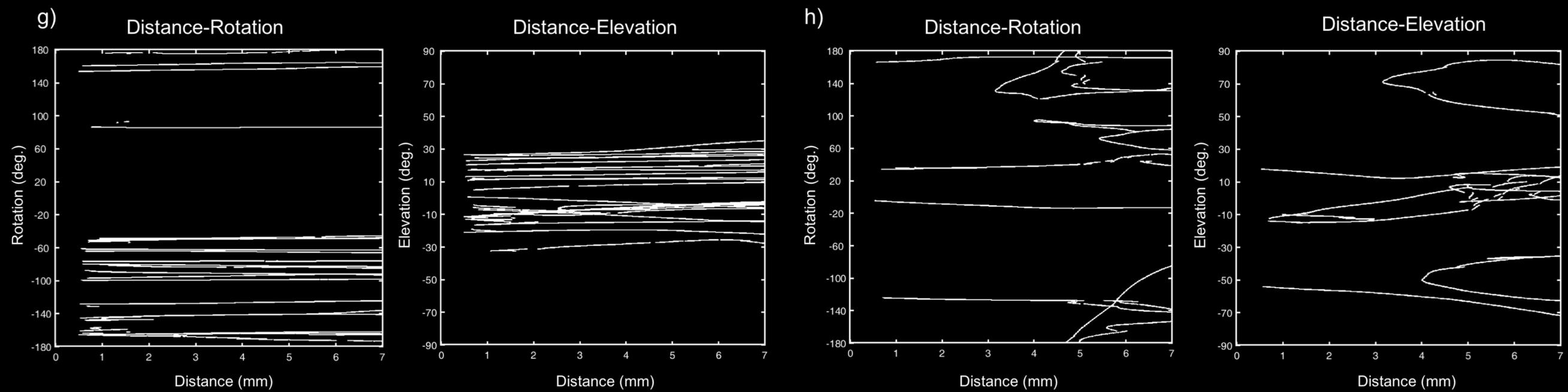

g) Distance-Rotation   Distance-Elevation    h) Distance-Rotation   Distance-Elevation

# Supplementary Data

*Supplementary Table 1. Composition of training and testing sets for each treatment cohort, including number of patients with available post-treatment response and survival information. pCR, pathologic complete response, MPR, major pathologic response, RECIST, Response Evaluation Criteria In Solid Tumors, RFS, recurrence-free survival, PFS, progression-free survival.*

|  | Training | | | | | Testing | | | | |
|---|---|---|---|---|---|---|---|---|---|---|
|  | | | *Response* | | *Prognosis* | | | *Response* | | *Prognosis* |
| Cohort | Notation | n | Endpoint | n | Endpoint | Notation | n | Endpoint | n | Endpoint |
| **BRCA-ACT** | $D_{tr}^1$ | 98 | pCR | 63 | RFS | $D_{te}^1$ | 144 | pCR | 94 | RFS |
| **BRCA-TCHP** | $D_{tr}^2$ | 69 | pCR | 0 | - | $D_{te}^2$ | 60 | pCR | 0 | - |
| **NSCLC-PLAT** | $D_{tr}^3$ | 53 | RECIST response | 53 | PFS | $D_{te}^3$ | 44 | RECIST response | 39 | PFS |
| **NSCLC-TRI** | $D_{tr}^4$ | 44 | MPR | 44 | RFS | $D_{te}^4$ | 46 | MPR | 46 | RFS |

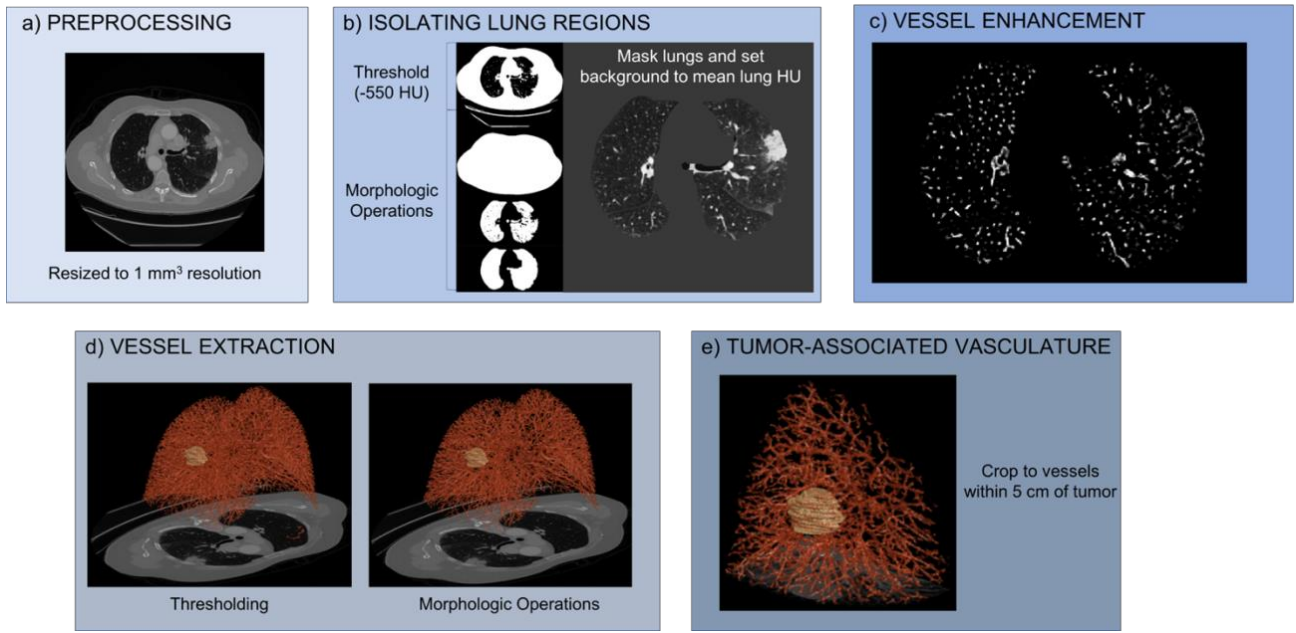

*Supplementary Figure 1. a) all volumes were resized to consistent isotropic resolution, b) a mask of the lung was obtained one slice at a time to isolate the lungs within a CT volume. First, a threshold of -550 HU was applied to identify non-lung, body tissue (mask 1). Hole filling and largest object detection were performed to obtain a mask of the full body (mask 2). The difference between mask 2 and 1 giving a raw lung segmentation (mask 3). Mask 3 was refined by morphologic operations such as opening, closing, and hole filling, to yield a final lung tissue mask. The average attenuation value within the lungs was computed and used to fill voxels corresponding to non-lung tissue so as not to create edges with shape attenuation changes at the lung interface. c) a vessel enhancement filter was applied to the lung volume to identify bright objects with tubular, vessel-like shapes (see Supplementary Implementation Details). d) Thresholding via Otsu's method was applied to the vessel-enhanced volume to isolate vessels from background (left). Morphologic processing, such as the removal of small and spherical objects, was applied to refine the vessel skeleton and remove noise (right). e) Finally, the volume is cropped to the tumor and its surrounding vasculature out to 5 cm in all directions.*

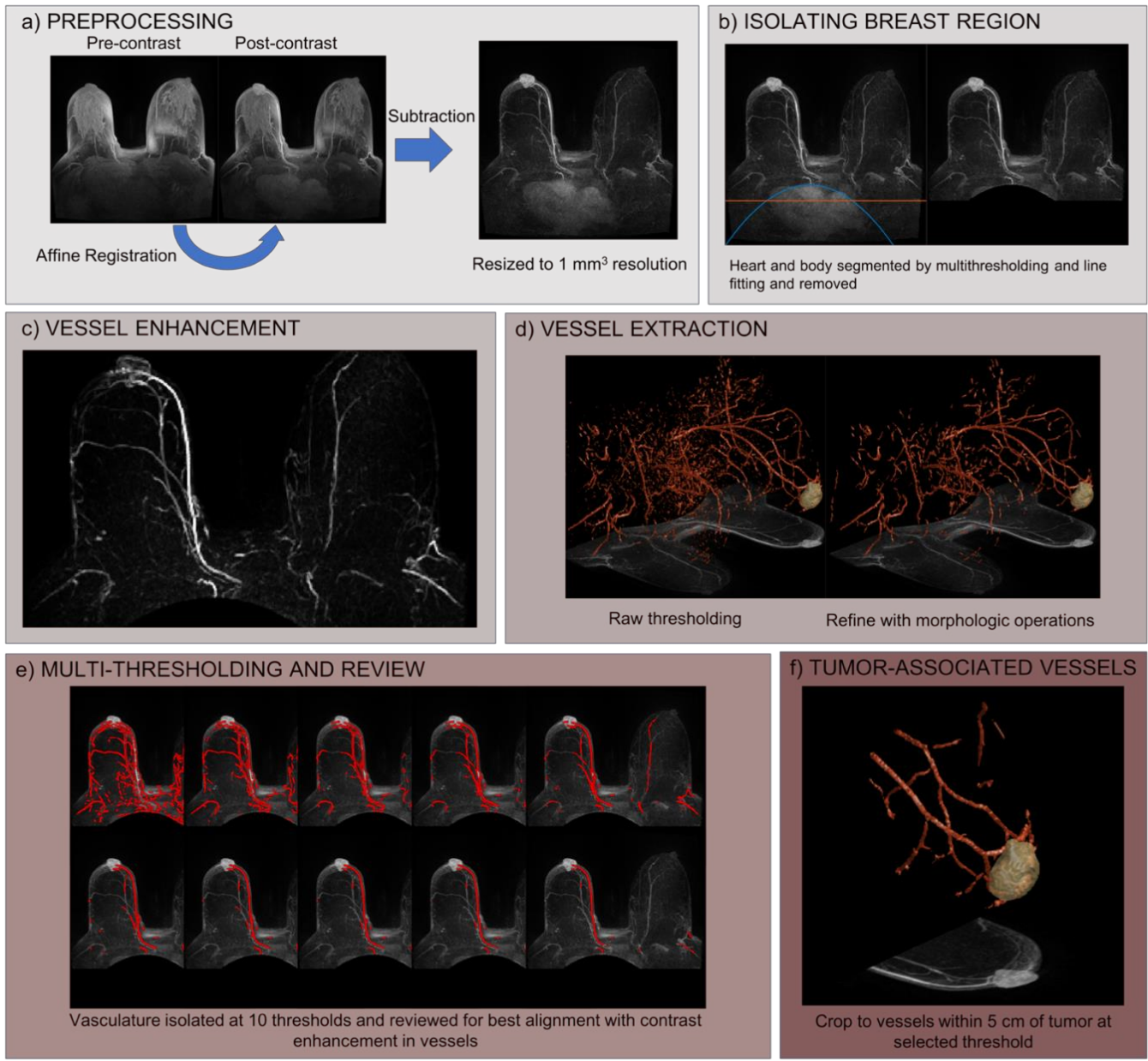

Supplementary Figure 2. a) An affine registration was performed to align the pre-contrast and first post-contrast DCE-MRI acquisitions, and the difference in image intensities was computed to yield subtraction volumes. Tissues perfused with contrast agent (such as the vessels and tumor) appear brightest, while much of the signal from surrounding breast tissue is removed. b) The heart was identified by multi-thresholding and largest object detection. Lines were fit to its upper edge (blue) and centroid (orange) and used to mask out contrast enhancement outside the breasts. c) a vessel enhancement filter was applied to the image with settings scaled relative to the magnitude of scan intensity (see Supplementary implementation details. d) Once a threshold was applied to isolate vessel voxels from background, the same morphologic operations applied in Supplementary Figure 1d were applied to remove artifacts. e) Thresholding was applied using Otsu's method at ten different thresholds, each yielding an increasingly sparse vessel segmentation. Due to the non-quantitative nature of MRI, the optimal threshold differed between scans and was determined through outcome-blinded manual review. f) Finally, volumes were cropped to the local tumor vasculature at a distance of 5 cm from the tumor in all directions.

*Supplementary Table 2. Full list of 61 QuanTAV Morphology features extracted.*

| Features | Description |
|---|---|
| Statistics of torsion per branch (f1-f5) | *Mean*, standard deviation (*std*), maximum (*max*), skewness (*skew*), and kurtosis (*kurt*) of torsion across all branches |
| Statistics of curvature standard deviation per branch (f6-f10) | *Mean, std, max, skew, kurt* of the standard deviation of curvature measured along each branch |
| Statistics of mean curvature per branch (f11-f15) | *Mean, std, max, skew, kurt* of the average curvature measured along each branch |
| Statistics of maximum curvature per branch (f16-f20) | *Mean, std, max, skew, kurt* of the maximum curvature measured along each branch |
| Statistics of curvature skewness per branch (f21-f25) | *Mean, std, max, skew, kurt* of the skewness of curvature measured along each branch |
| Statistics of curvature kurtosis per branch (f26-f30) | *Mean, std, max, skew, kurt* of the kurtosis of curvature measured along each branch |
| Statistics of global vascular curvature (f31-f35) | *Mean, std, max, skew, kurt* of the curvature measured across all branches combined |
| Histogram of global vascular curvature (f36-f45) | 10-bin histogram of the curvature measured across all points of the vessel volume |
| Histogram of torsion (f46-f55) | 10-bin histogram of the torsion measured across all branches combined |
| Total vessel volume (f56-f58) | Vessel volume (f56), vessel volume normalized to the total size of the 3D region of interest (f57), vessel volume normalized to the volume of the tumor (f58). |
| Total vessel length (f59) | Total length of vessels within the region of interest |
| Tumor feeding branches (f60, f61) | Number (f60) and percentage (f61) of vessel branches that enter the tumor volume from the surrounding tumor environment. |

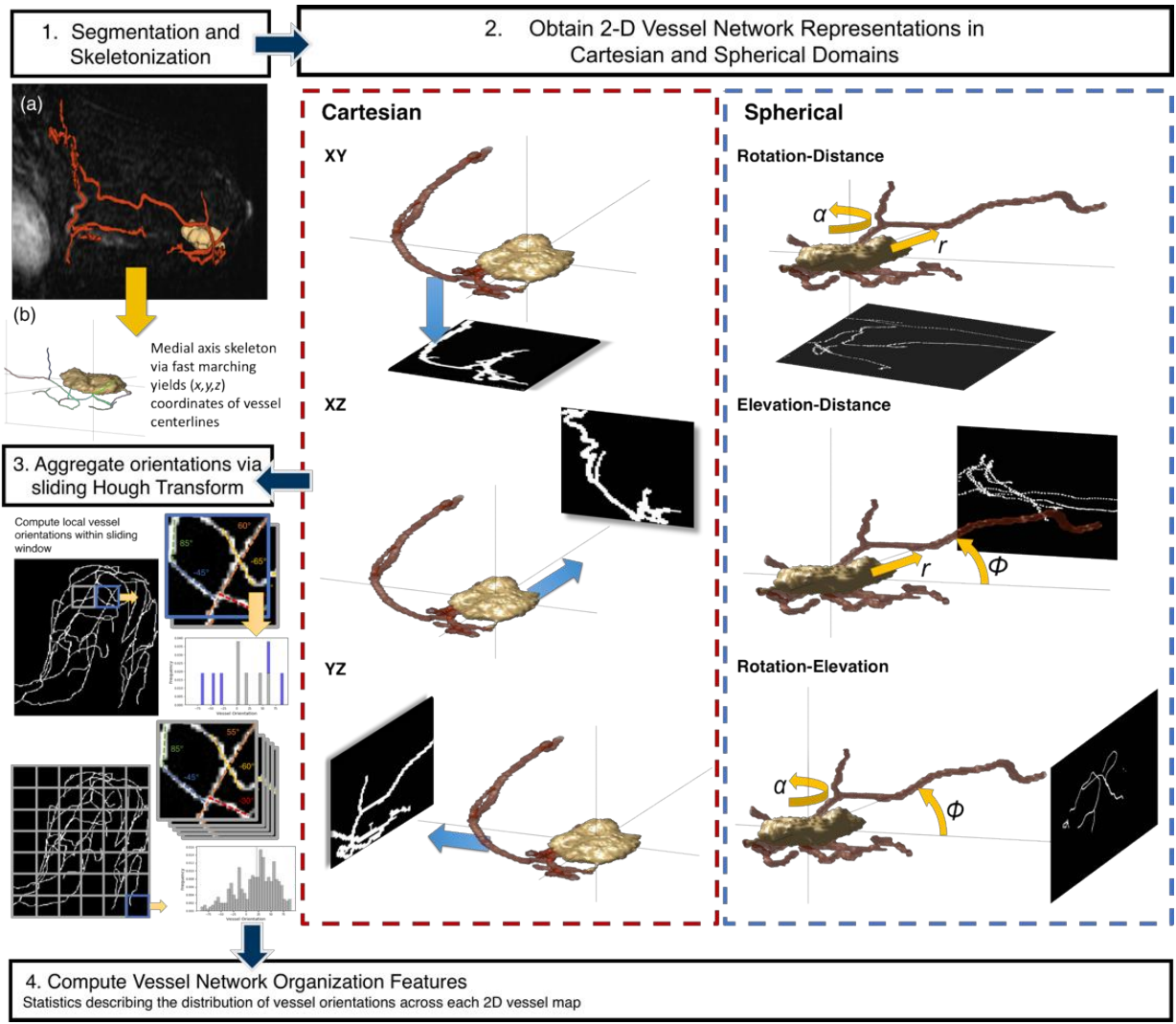

*Supplementary Figure 3. Workflow for computing features describing tumor-associated vasculature organization. QuanTAV Spatial Organization[29] features quantify the distribution of local vessels within a fixed radius surrounding the tumor by creating 2D projection images of a vessel's position in cartesian (X,Y,Z) space and spherical (rotation and elevation relative to the tumor surface, and distance relative to the tumor surface). Each project image is then analyzed locally within a sliding window. The Hough transform[93] is applied to detect lines within the window and quantify their orientation. The most prominent vessel orientations, up to a maximum of five, are stored. Statistics of the distribution of vessel orientations form the set of QuanTAV spatial organization features. The maximum distance from the tumor and size of the sliding window are optimized for each cancer domain/imaging modality by performance of a classifier in cross-validation in the training sets.*

Supplementary Table 3. Full list of 30 QuanTAV Spatial Organization features extracted.

| Features | Description |
|---|---|
| Statistics of vessel orientation along XY projection image (f1-f5) | Mean, median (med), standard deviation (std), skewness (skew), and kurtosis (kurt) of local vessel orientations computed across XY vessel map |
| Statistics of vessel orientation along the XZ projection image (f6-f10) | Mean, med, std, skew, kurt of local vessel orientations computed across XZ vessel map |
| Statistics of vessel orientation along the YZ projection image (f11-f15) | Mean, std, max, skew, kurt of local vessel orientations computed across XZ vessel map |
| Statistics of vessel orientation along the rotation-elevation projection image (f16-f20) | Mean, std, max, skew, kurt of local vessel orientations computed across vessel map of rotation and elevation with respect to the tumor |
| Statistics of vessel orientation along the distance-rotation projection image (21-f25) | Mean, std, max, skew, kurt of local vessel orientations computed across vessel map of distance and rotation with respect to the tumor |
| Statistics of vessel orientation along the distance-elevation projection image (f26-f30) | Mean, std, max, skew, kurt of local vessel orientations computed across vessel map of distance and elevation with respect to the tumor |

Supplementary Table 4. Top features and corresponding coefficients for LDA classifier to predict pathologic response in HER2-negative breast cancer patients receiving BRCA-ACT. Expression of features with positive coefficients contributes to a response prediction, while expression of features with negative coefficients contributes to a prediction of non-response.

| Feature Name | Coefficient |
|---|---|
| QuanTAV Spatial Organization - YZ - Skewness | 0.18 |
| QuanTAV Spatial Organization - Distance-Elevation - Median | 0.72 |
| QuanTAV Spatial Organization - Distance-Elevation - Mean | -0.16 |
| QuanTAV Spatial Organization - XZ - Skewness | 0.060 |
| QuanTAV Spatial Organization - YZ - Median | 0.88 |
| QuanTAV - Torsion - Mean | -0.44 |
| Constant | -0.23 |

Supplementary Table 5. Area under the receiver operating characteristic curve (AUC) of QuanTAV response score models and significance.

| Cohort | Training | Testing | |
|---|---|---|---|
| | AUC (95% CI) | AUC (95% CI) | p-value |
| BRCA-ACT | 0.63 (0.55-0.70) | 0.65 (0.54-0.76) | 0.009 |
| BRCA-TCHP | 0.65 (0.58-0.72) | 0.63 (0.47-0.76) | 0.042 |
| NSCLC-PLAT | 0.68 (0.56-0.79) | 0.70 (0.54-0.85) | 0.024 |
| NSCLC-TRI | 0.71 (0.60-0.81) | 0.71 (0.51-0.84) | 0.0093 |

Supplementary Table 6. Univariate (UVA) and multivariable (MVA) analysis of QuanTAV response score and available clinical variables for prediction of pathologic response to BRCA-ACT. Clinical variables that were individually significant in the training set (hormone receptor status) were incorporated into logistic regression models alone and with QuanTAV response score, and evaluated on the testing set. OR, odds ratio; p, p-value.

| | Training set | (n=98) | | | Testing set | (n=144) | | |
|---|---|---|---|---|---|---|---|---|
| | *Univariable* | | *Multivariable* | | *Univariable* | | *Multivariable* | |
| **Variable** | Odds Ratio (95% CI) | p | Odds Ratio (95% CI) | p | Odds Ratio (95% CI) | p | Odds Ratio (95% CI) | p |
| QuanTAV Response Score | 0.09 (0.01 - 0.90) | **0.046** | 0.05 (0.00 - 0.87) | **0.040** | 0.07 (0.01 - 0.63) | **0.018** | 0.02 (0.00 - 0.32) | **0.005** |
| Hormone Receptor Status (Positive vs. Negative) | 4.73 (1.77 - 12.65) | **0.002** | 5.99 (2.01 - 17.79) | **0.001** | 5.10 (2.02 - 12.90) | **0.001** | 7.24 (2.49 - 21.05) | **<1E-5** |
| Age (per year increase) | 1.02 (0.97 - 1.07) | 0.495 | 1.06 (0.99 - 1.12) | 0.072 | 1.06 (1.01 - 1.11) | **0.012** | 1.06 (1.01 - 1.12) | **0.019** |
| Lesion Diameter (per mm increase) | 1.17 (0.96 - 1.41) | 0.115 | 1.23 (0.98 - 1.53) | 0.069 | 1.09 (0.93 - 1.28) | 0.268 | 1.05 (0.88 - 1.26) | 0.559 |

Supplementary Table 7. The set of features selected by the elastic-net Cox regression model as being most prognostic of RFS from initiation of neoadjuvant chemotherapy among breast cancer patients receiving BRCA-ACT and their corresponding hazard ratios. The hazard ratios shown here reflect the risk associated with an increase of one standard deviation in feature value on the training set. A hazard ratio of less than 1 implies that an increase in that feature's value is associated with reduced risk, while a hazard ratio greater than 1 implies the opposite.

| **Feature Name** | **Hazard Ratio** |
|---|---|
| QuanTAV Spatial Organization - Distance-Rotation - Skewness | 0.89 |
| QuanTAV Spatial Organization - Distance-Elevation - Standard Deviation | 0.89 |
| QuanTAV Spatial Organization - Rotation-Elevation - Skewness | 0.90 |
| QuanTAV Spatial Organization - Distance-Rotation - Standard Deviation | 0.93 |
| QuanTAV Spatial Organization - XZ - Kurtosis | 0.93 |
| Ratio of Vessel to Tumor Volume | 0.96 |
| QuanTAV Morphology - Torsion Histogram - Bin 2 | 0.98 |
| QuanTAV Spatial Organization - YZ - Skewness | 1.00 |
| QuanTAV Spatial Organization - YZ - Kurtosis | 1.03 |
| QuanTAV Morphology - Torsion - Mean | 1.03 |
| No. Vessels Feeding Tumor | 1.05 |
| QuanTAV Morphology - Torsion - Standard Deviation | 1.06 |
| QuanTAV Spatial Organization - Rotation-Elevation - Kurtosis | 1.08 |
| Percentage of Vessels Feeding Tumor | 1.11 |

Supplementary Table 8. Hazard ratio (HR), concordance index (C-index), and p-value of the HR for each prognostic model in the training and testing sets.

| **Cohort** | **Signature** | Training | | | Testing | | |
|---|---|---|---|---|---|---|---|
| | | *Hazard Ratio (95% CI)* | *C-index* | *p* | *Hazard Ratio (95% CI)* | *C-index* | *p* |
| **BRCA-ACT** | *Risk Score* | 1.43 (1.24-1.66) | 0.79 | **<1e-5** | 1.25 (1.08-1.44) | 0.66 | **0.002** |
| | *Risk Group* | 10.75 (1.43-80.61) | 0.66 | **0.021** | 4.25 (1.29-14.07) | 0.62 | **0.018** |
| **NSCLC-PLAT** | *Risk Score* | 1.32 (1.16-1.50) | 0.77 | **1.6e-5** | 1.12 (0.96-1.31) | 0.61 | 0.14 |
| | *Risk Group* | 7.19 (2.85-18.14) | 0.71 | **3.0e-5** | 2.29 (1.07-4.94) | 0.62 | **0.034** |
| **NSCLC-TRI** | *Risk Score* | 1.32 (1.16-1.50) | 0.81 | **1.9e-5** | 1.28 (1.01-1.62) | 0.66 | **0.039** |
| | *Risk Group* | 20.33 (2.68-154.12) | 0.74 | **4.5e-5** | 3.77 (1.09-13.00) | 0.64 | **0.036** |

*Supplementary Table 9. Cox proportional hazard univariable (UVA) and multivariable (MVA) analysis of recurrence free survival following BRCA-ACT treatment, including QuanTAV risk score, QuanTAV risk groups, baseline clinical variables, and post-chemotherapy response.*

| Variable | Univariable | | Multivariable Risk Score (Continuous) | | Multivariable Risk Groups (Categorical) | |
|---|---|---|---|---|---|---|
| | Hazard Ratio (95% CI) | p | Hazard Ratio (95% CI) | p | Hazard Ratio (95% CI) | p |
| *QuanTAV Risk Score (increase of 1)* | 1.25 (1.08 - 1.44) | **0.002** | 1.20 (1.04 - 1.40) | **0.014** | -- | -- |
| *QuanTAV Risk Group (High vs. Low Risk)* | 4.25 (1.29 - 14.07) | **0.018** | -- | -- | 5.51 (1.41 - 21.49) | **0.014** |
| *Hormone receptor status (positive vs. negative)* | 0.45 (0.22 - 0.95) | **0.036** | 0.40 (0.18 - 0.89) | **0.025** | 0.36 (0.16 - 0.81) | **0.014** |
| *Age (per year increase)* | 0.95 (0.92 - 0.99) | **0.017** | 0.94 (0.90 - 0.99) | **0.013** | 0.94 (0.90 - 0.98) | **0.007** |
| *Largest lesion diameter (per mm increase)* | 1.10 (1.00 - 1.22) | 0.062 | 1.01 (0.90 - 1.14) | 0.821 | 1.02 (0.90 - 1.15) | 0.746 |
| *Functional Tumor Volume (per 10cc increase)* | 1.21 (1.07 - 1.36) | **0.002** | 1.13 (1.01 - 1.28) | **0.040** | 1.16 (1.04 - 1.31) | **0.011** |

*Supplementary Table 10. Correlation of features in BRCA-ACT risk score associated with functional tumor volume (FTV) on DCE-MRI.*

| Feature Name | Correlation Coefficient | P-value |
|---|---|---|
| *QuanTAV Spatial Organization - XZ - Kurtosis* | 0.0253 | 0.8086 |
| *QuanTAV Spatial Organization - YZ - Skewness* | -0.0392 | 0.7074 |
| *QuanTAV Spatial Organization - YZ - Kurtosis* | -0.1246 | 0.2315 |
| *QuanTAV Spatial Organization - Rotation-Elevation - Skewness* | 0.0006 | 0.9954 |
| *QuanTAV Spatial Organization - Rotation-Elevation - Kurtosis* | -0.1103 | 0.2900 |
| *QuanTAV Spatial Organization - Distance-Rotation - Standard Deviation* | -0.0843 | 0.4192 |
| *QuanTAV Spatial Organization - Distance-Rotation - Skewness* | -0.2126 | **0.0396** |
| *QuanTAV Spatial Organization - Distance-Elevation - Standard Deviation* | -0.2115 | **0.0407** |
| *QuanTAV Morphology - Torsion - Mean* | 0.0946 | 0.3645 |
| *QuanTAV Morphology - Torsion - Standard Deviation* | 0.1237 | 0.2348 |
| *Ratio of Vessel to Tumor Volume* | -0.1905 | 0.0660 |
| *QuanTAV Morphology - Torsion Histogram - Bin 2* | -0.1549 | 0.1361 |
| *No. Vessels Feeding Tumor* | 0.2294 | **0.0262** |
| *Percentage of Vessels Feeding Tumor* | 0.5515 | **<1E-5** |

*Supplementary Table 11. Top features and corresponding coefficients for LDA classifier to predict pathologic response in HER2-positive breast cancer patients receiving HER2-targeted neoadjuvant chemotherapy (BRCA-TCHP). Expression of features with positive coefficients contributes to a response prediction, while expression of features with negative coefficients contributes to a prediction of non-response.*

| Feature Name | Coefficient |
|---|---|
| *QuanTAV Spatial Organization - XY - Skewness* | -0.52 |
| *QuanTAV Spatial Organization - Distance-Elevation - Mean* | 0.30 |
| *QuanTAV Spatial Organization - Distance-Elevation - Median* | 0.48 |
| *QuanTAV Morphology - Torsion Histogram - Bin 7* | -0.50 |
| *Constant* | 0.029 |

*Supplementary Table 12. Univariate (UVA) and multivariable (MVA) analysis of QuanTAV response score and available clinical variables for prediction of pathologic response to BRCA-TCHP. Clinical variables that were individually significant in the training set (hormone receptor status) were incorporated into logistic regression models alone and with QuanTAV response score, and evaluated on the testing set. OR, odds ratio; p, p-value.*

| | Training set (n=69) | | | | Testing set (n=60) | | | |
|---|---|---|---|---|---|---|---|---|
| | *Univariable* | | *Multivariable* | | *Univariable* | | *Multivariable* | |
| **Variable** | Odds Ratio (95% CI) | p | Odds Ratio (95% CI) | p | Odds Ratio (95% CI) | p | Odds Ratio (95% CI) | p |
| *QuanTAV Response Score* | 0.06 (0.00 - 0.69) | **0.024** | 0.13 (0.01 - 1.88) | 0.135 | 0.17 (0.01 - 1.88) | 0.148 | 0.17 (0.01 - 2.38) | 0.188 |
| *Hormone receptor status (positive vs. negative)* | 3.68 (1.23 - 11.05) | **0.020** | 3.11 (0.91 - 10.68) | 0.072 | 3.89 (1.23 - 12.29) | **0.021** | 3.86 (1.17 - 12.77) | **0.027** |
| *Age (per year increase)* | 1.02 (0.97 - 1.06) | 0.475 | 1.02 (0.98 - 1.07) | 0.343 | 0.99 (0.94 - 1.04) | 0.739 | 1.00 (0.95 - 1.06) | 0.865 |
| *Largest lesion diameter (per mm increase)* | 1.03 (0.89 - 1.19) | 0.730 | 1.01 (0.85 - 1.20) | 0.925 | 1.00 (0.82 - 1.23) | 0.993 | 0.99 (0.77 - 1.27) | 0.934 |
| *Clinical Stage (per stage increase)* | 1.01 (0.46 - 2.18) | 0.986 | 0.74 (0.29 - 1.88) | 0.533 | 0.97 (0.33 - 2.83) | 0.956 | 1.03 (0.27 - 3.92) | 0.964 |

Supplementary Table 13. Top features and corresponding coefficients for LDA classifier to predict RECIST response in NSCLC patients receiving platinum-based neoadjuvant chemotherapy (NSCLC-PLAT). Expression of features with positive coefficients contributes to a response prediction, while expression of features with negative coefficients contributes to a prediction of non-response.

| Feature Name | Coefficient |
| --- | --- |
| *QuanTAV Spatial Organization - XZ - Kurtosis* | 0.86 |
| *QuanTAV Spatial Organization - XZ - Skewness* | -0.53 |
| *QuanTAV Spatial Organization - Distance-Rotation - Skewness* | -0.44 |
| *QuanTAV Spatial Organization - Distance-Elevation - Standard Deviation* | -0.98 |
| *QuanTAV Spatial Organization - XZ - Median* | -1.02 |
| *QuanTAV Spatial Organization - Distance-Rotation - Mean* | -0.22 |
| *Constant* | -0.015 |

Supplementary Table 14. Univariate (UVA) and multivariable (MVA) analysis of QuanTAV response score and available clinical variables for prediction of RECIST response in NSCLC-PLAT recipients. Clinical variables that were individually significant in the training set (age) were incorporated into logistic regression models alone and with QuanTAV response score, and evaluated on the testing set. OR, odds ratio; p, p-value.

| | **Training set** | (n=53) | | | **Testing set** | (n=44) | | |
| --- | --- | --- | --- | --- | --- | --- | --- | --- |
| | *Univariable* | | *Multivariable* | | *Univariable* | | *Multivariable* | |
| **Variable** | Odds Ratio (95% CI) | p | Odds Ratio (95% CI) | p | Odds Ratio (95% CI) | p | Odds Ratio (95% CI) | p |
| *QuanTAV Response Score* | 0.04 (0.00 - 0.45) | **0.010** | 0.03 (0.00 - 0.55) | **0.017** | 0.04 (0.00 - 0.62) | **0.021** | 0.03 (0.00 - 0.93) | **0.045** |
| *Age (per year increase)* | 0.94 (0.89 - 1.00) | **0.048** | 0.95 (0.89 - 1.03) | 0.228 | 1.03 (0.98 - 1.07) | 0.232 | 1.03 (0.98 - 1.09) | 0.236 |
| *Sex (male vs. female)* | 0.37 (0.12 - 1.12) | 0.078 | 0.35 (0.09 - 1.42) | 0.141 | 0.93 (0.27 - 3.17) | 0.902 | 0.87 (0.20 - 3.79) | 0.856 |
| *Stage (per stage increase)* | 1.02 (0.50 - 2.10) | 0.950 | 1.05 (0.38 - 2.91) | 0.924 | 1.44 (0.55 - 3.78) | 0.458 | 1.06 (0.30 - 3.72) | 0.925 |
| *Longest diameter (per mm increase)* | 0.99 (0.97 - 1.01) | 0.534 | 0.99 (0.96 - 1.03) | 0.641 | 1.01 (0.99 - 1.02) | 0.454 | 1.00 (0.98 - 1.03) | 0.670 |
| *Histology (SCC/other vs. Adenocarcinoma)* | 0.24 (0.04 - 1.28) | 0.094 | 0.25 (0.03 - 2.24) | 0.213 | 1.02 (0.29 - 3.65) | 0.975 | 1.97 (0.40 - 9.62) | 0.404 |
| *Former smoker (Yes vs. No)* | 3.75 (0.68 - 20.63) | 0.129 | 7.11 (0.61 - 82.78) | 0.118 | 0.40 (0.07 - 2.35) | 0.311 | 0.29 (0.04 - 2.16) | 0.226 |

Supplementary Table 15. The set of features selected by the elastic-net Cox regression model as being most prognostic of PFS following inception of platinum-based chemotherapy (NSCLC-PLAT) among lung cancer patients and corresponding hazard ratios in the training set. The hazard ratios shown here reflect the risk of an increase of one standard deviation in feature value on the training set. A hazard ratio of less than 1 implies that an increase in that feature's value is associated with reduced risk, while a hazard ratio greater than 1 implies the opposite.

| Feature Name | Hazard Ratio |
| --- | --- |
| Normalized Vessel Volume | 0.48 |
| QuanTAV Morphology - Global Curvature - Kurtosis | 0.61 |
| QuanTAV Spatial Organization - XZ - Skewness | 0.73 |
| QuanTAV Morphology - Kurtosis of Curvature Per Vessel - Standard Deviation | 0.75 |
| QuanTAV Morphology - Global Curvature - Skewness | 0.86 |
| QuanTAV Morphology - Maximum Curvature per Vessel - Mean | 0.88 |
| QuanTAV Morphology - Skewness of Curvature Per Vessel - Kurtosis | 0.90 |
| QuanTAV Morphology - Skewness of Curvature Per Vessel - Skewness | 0.97 |
| QuanTAV Morphology - Torsion Histogram - Bin 8 | 1.07 |
| QuanTAV Spatial Organization - Distance-Elevation - Mean | 1.16 |
| QuanTAV Spatial Organization - Distance-Rotation - Kurtosis | 1.28 |
| QuanTAV Spatial Organization - Distance-Elevation - Standard Deviation | 1.80 |
| QuanTAV Spatial Organization - Distance-Rotation - Standard Deviation | 2.24 |
| QuanTAV Spatial Organization - XZ - Median | 2.70 |
| QuanTAV Spatial Organization - Distance-Elevation - Kurtosis | 2.86 |

Supplementary Table 16. Cox proportional hazard univariable (UVA) and multivariable (MVA) analysis of 10-year progression free survival following NSCLC-PLAT treatment, including QuanTAV risk score, QuanTAV risk groups, baseline clinical variables, and post-chemotherapy response.

| Variable | Univariable Hazard Ratio (95% CI) | p | Multivariable Risk Score (Continuous) Hazard Ratio (95% CI) | p | Multivariable Risk Groups (Categorical) Hazard Ratio (95% CI) | p |
| --- | --- | --- | --- | --- | --- | --- |
| QuanTAV Risk Score (increase of 1) | 1.12 (0.96 - 1.31) | 0.141 | 1.10 (0.93 - 1.31) | 0.260 | -- | -- |
| QuanTAV Risk Group (High vs. Low Risk) | 2.29 (1.07 - 4.94) | **0.034** | -- | -- | 2.53 (1.10 - 5.80) | **0.028** |
| Age (per year increase) | 0.99 (0.96 - 1.02) | 0.657 | 1.00 (0.97 - 1.03) | 0.830 | 1.00 (0.97 - 1.03) | 0.934 |
| Sex (male vs. female) | 1.06 (0.52 - 2.18) | 0.868 | 1.28 (0.56 - 2.96) | 0.560 | 1.43 (0.63 - 3.27) | 0.389 |
| Stage (per stage increase) | 1.43 (0.79 – 2.58) | 0.241 | 1.41 (0.68 – 2.93) | 0.360 | 1.60 (0.72 – 3.54) | 0.248 |
| Longest diameter (per mm increase) | 1.00 (0.99 - 1.01) | 0.543 | 1.00 (0.99 - 1.01) | 0.692 | 1.00 (0.99 - 1.02) | 0.436 |
| Histology (Adenocarcinoma vs. SCC/other) | 0.82 (0.38 - 1.77) | 0.619 | 0.92 (0.42 - 2.01) | 0.827 | 0.91 (0.42 - 1.97) | 0.807 |
| Former smoker | 0.70 (0.30 - 1.65) | 0.419 | 0.72 (0.29 - 1.80) | 0.477 | 0.76 (0.30 - 1.91) | 0.565 |

Supplementary Table 17. Top features and corresponding coefficients for LDA classifier to predict pathologic response in NSCLC patients receiving neoadjuvant chemoradiation (NSCLC-TRI). Expression of features with positive coefficients contributes to a response prediction, while expression of features with negative coefficients contributes to a prediction of non-response.

| Feature Name | Coefficient |
| --- | --- |
| *QuanTAV Spatial Organization - Distance-Elevation - Kurtosis* | -0.89 |
| *QuanTAV Morphology - Skewness of Curvature Per Vessel - Kurtosis* | 1.21 |
| *QuanTAV Spatial Organization - Distance-Rotation - Mean* | 0.71 |
| *QuanTAV Morphology - Torsion - Max* | 0.41 |
| *Constant* | -0.14 |

Supplementary Table 18. Univariate (UVA) and multivariable (MVA) analysis of QuanTAV response score and available clinical variables for prediction of pathologic response in NSCLC-TRI recipients. Clinical variables that were individually significant in the training set (histology) were incorporated into logistic regression models alone and with QuanTAV response score, and evaluated on the testing set. OR, odds ratio; p, p-value.

| | **Training set** | (n=44) | | | **Testing set** | (n=46) | | |
| --- | --- | --- | --- | --- | --- | --- | --- | --- |
| | *Univariable* | | *Multivariable* | | *Univariable* | | *Multivariable* | |
| **Variable** | Odds Ratio (95% CI) | p | Odds Ratio (95% CI) | p | Odds Ratio (95% CI) | p | Odds Ratio (95% CI) | p |
| *QuanTAV Response Score* | 0.02 (0.00-0.25) | **0.003** | 0.00 (0.00-0.25) | **0.017** | 0.06 (0.00-0.64) | **0.020** | 0.00 (0.00-0.78) | **0.042** |
| *Age (per year increase)* | 1.00 (0.94-1.07) | 0.911 | 1.04 (0.92-1.19) | 0.510 | 1.03 (0.98-1.09) | 0.283 | 0.95 (0.84-1.06) | 0.338 |
| *Sex (male vs. female)* | 3.25 (0.93-11.41) | 0.066 | 21.48 (0.68-675.21) | 0.081 | 2.00 (0.59-6.83) | 0.269 | 1.04 (0.11-9.94) | 0.972 |
| *Stage (IIIA vs IIIB)* | N/A | N/A | N/A | N/A | 0.13 (0.01-1.27) | 0.080 | 0.15 (0.01-3.51) | 0.238 |
| *Longest diameter (per mm increase)* | 0.99 (0.97-1.01) | 0.160 | 0.97 (0.90-1.04) | 0.407 | 0.95 (0.92-0.98) | **0.004** | 0.97 (0.93-1.02) | 0.208 |
| *Histology (Adenocarcinoma vs. SCC/other)* | 8.36 (1.52-46.15) | **0.015** | 35.61 (1.37-928.63) | **0.032** | 9.43 (2.28-39.04) | **0.002** | 15.47 (1.40-171.31) | **0.026** |
| *ECOG performance status (per grade increase)* | 3.06 (0.74-12.65) | 0.122 | 7.45 (0.40-137.06) | 0.177 | 1.00 (0.24-4.20) | 1.000 | 0.41 (0.02-10.16) | 0.583 |
| *Chemotherapy (Carboplatin vs. Cisplatin)* | 1.53 (0.37-6.35) | 0.557 | 5.44 (0.15-193.86) | 0.353 | 0.19 (0.05-0.78) | **0.021** | 0.11 (0.00-3.33) | 0.205 |
| *Radiotherapy Induction Dose (per Gy increase)* | 0.98 (0.92-1.04) | 0.560 | 1.00 (0.89-1.12) | 0.960 | 0.95 (0.89-1.01) | 0.123 | 0.90 (0.78-1.04) | 0.140 |

Supplementary Table 19. The set of features selected by the elastic-net Cox regression model as being most prognostic of RFS from date of surgery among lung cancer patients receiving trimodality therapy and corresponding hazard ratios in the training set. The hazard ratios shown here reflect the risk of an increase of one standard deviation in feature value on the training set. A hazard ratio of less than 1 implies that an increase in that feature's value is associated with reduced risk, while a hazard ratio greater than 1 implies the opposite.

| Feature Name | Hazard Ratio |
|---|---|
| *QuanTAV Morphology - Skewness of Curvature Per Vessel - Kurtosis* | 0.71 |
| *QuanTAV Spatial Organization - YZ - Skewness* | 0.77 |
| *QuanTAV Morphology - Torsion Histogram - Bin 9* | 0.82 |
| *QuanTAV Spatial Organization - Distance-Rotation - Median* | 0.90 |
| *QuanTAV Spatial Organization - XY - Standard Deviation* | 0.92 |
| *QuanTAV Spatial Organization - Rotation-Elevation - Skewness* | 0.93 |
| *QuanTAV Spatial Organization - YZ - Kurtosis* | 0.98 |
| *QuanTAV Morphology - Torsion Histogram - Bin 10* | 0.99 |
| *QuanTAV Spatial Organization - XY - Mean* | 0.99 |
| *QuanTAV Morphology - Curvature Histogram - Bin 6* | 1.01 |
| *QuanTAV Morphology - Maximum Curvature per Vessel - Mean* | 1.05 |
| *QuanTAV Spatial Organization - Distance-Rotation - Kurtosis* | 1.07 |
| *QuanTAV Morphology - Skewness of Curvature Per Vessel - Mean* | 1.08 |
| *QuanTAV Spatial Organization - Distance-Rotation - Standard Deviation* | 1.24 |
| *QuanTAV Morphology - Deviation of Curvature Per Vessel - Standard Deviation* | 1.31 |

Supplementary Table 20. Cox proportional hazard univariable (UVA) and multivariable (MVA) analysis of recurrence free survival following NSCLC-TRI treatment, including QuanTAV risk score, QuanTAV risk groups, baseline clinical variables, and post-chemotherapy response and histolopathologic variables.

| Variable | Univariable | | Multivariable Risk Score (Continuous) | | Multivariable Risk Groups (Categorical) | |
|---|---|---|---|---|---|---|
| | Hazard Ratio (95% CI) | p | Hazard Ratio (95% CI) | p | Hazard Ratio (95% CI) | p |
| QuanTAV Risk Score (increase of 1) | 1.28 (1.01-1.62) | **0.039** | 1.76 (1.16-2.67) | **0.008** | -- | -- |
| QuanTAV Risk Group (High vs. Low Risk) | 3.77 (1.09-13.00) | **0.036** | -- | -- | 31.91 (3.66-278.12) | **0.002** |
| Age (per year increase) | 1.03 (0.98-1.07) | 0.240 | 0.99 (0.92-1.07) | 0.862 | 0.97 (0.90-1.05) | 0.460 |
| Sex (male vs. female) | 1.16 (0.48-2.82) | 0.738 | 0.46 (0.11-1.83) | 0.270 | 0.52 (0.14-1.89) | 0.320 |
| Stage (IIIB vs IIIA) | 1.17 (0.34-3.99) | 0.807 | 0.79 (0.12-5.10) | 0.803 | 0.20 (0.03-1.56) | 0.126 |
| Longest diameter (per mm increase) | 1.00 (0.98-1.01) | 0.684 | 1.02 (0.99-1.05) | 0.145 | 1.05 (1.01-1.08) | **0.009** |
| Histology (Adenocarcinoma vs. SCC/other) | 2.18 (0.79-6.02) | 0.134 | 2.73 (0.54-13.89) | 0.226 | 2.94 (0.57-15.05) | 0.196 |
| ECOG performance status (per grade increase) | 0.86 (0.31-2.37) | 0.767 | 1.78 (0.36-8.73) | 0.477 | 1.01 (0.24-4.16) | 0.990 |
| Chemotherapy regimen (Cisplatin vs. Carboplatin) | 0.39 (0.11-1.34) | 0.135 | 0.19 (0.02-1.71) | 0.140 | 0.11 (0.01-1.29) | 0.079 |
| Radiotherapy Induction Dose (per Gy increase) | 0.98 (0.93-1.04) | 0.537 | 1.11 (1.01-1.22) | **0.035** | 1.14 (1.03-1.26) | **0.009** |
| Surgical procedure (pneumonectomy vs. lobectomy) | 0.48 (0.16-1.44) | 0.192 | 0.17 (0.03-1.05) | 0.056 | 0.03 (0.00-0.30) | **0.003** |
| Presence of vascular invasion | 3.56 (1.34-9.47) | **0.011** | 5.31 (0.86-32.65) | 0.071 | 6.45 (1.16-35.68) | **0.033** |
| Presence of lymphatic invasion | 2.42 (0.98-6.00) | 0.056 | 0.84 (0.17-4.07) | 0.831 | 1.80 (0.42-7.75) | 0.429 |

*Supplementary Table 21. Pearson's correlation of top 5 most prognostic QuanTAV features and QuanTAV risk score with risk score derived from intra- and peri-tumoral texture features within the NSCLC-TRI cohort.*

| Feature | Correlation Coefficient | p-value |
|---|---|---|
| QuanTAV Morphology - Torsion Histogram - Bin 9 | 0.070 | 0.51 |
| QuanTAV Spatial Organization - Distance-Rotation - Standard Deviation | -0.053 | 0.63 |
| QuanTAV Spatial Organization - YZ - Skewness | -0.044 | 0.68 |
| QuanTAV Morphology - Deviation of Curvature Per Vessel - Standard Deviation | 0.0159 | 0.88 |
| QuanTAV Morphology - Skewness of Curvature Per Vessel - Kurtosis | -0.408 | **6.5E-05** |
| QuanTAV Prognostic Risk Score | 0.229 | **0.030** |

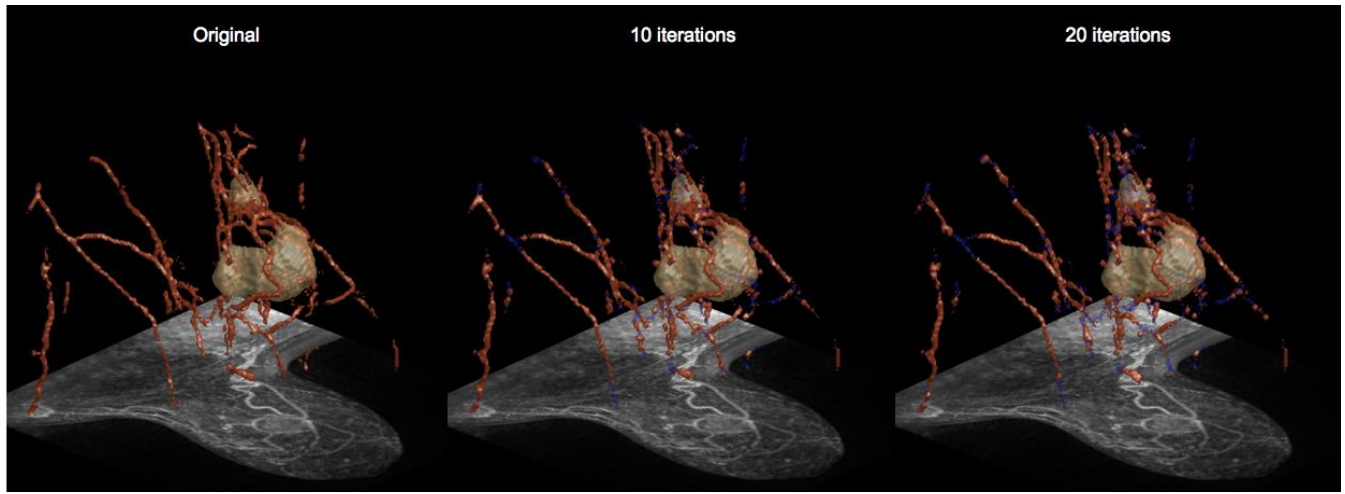
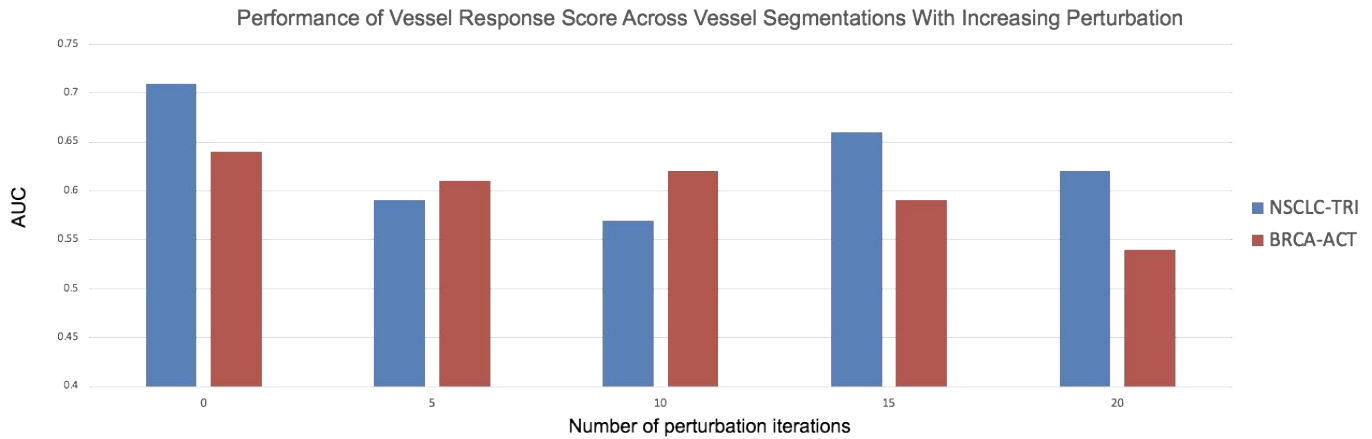

Supplementary Figure 4. Evaluating the robustness of QuanTAV-based response prediction to errors in vessel segmentation within a breast MRI (BRCA-ACT, n=144) and lung CT (NSCLC-TRI, n=46) testing set. Top: Vessel segmentations were randomly eroded and dilated at branchpoints and endpoints for increasing numbers of iterations. Vessel voxels in red were retained in the vasculature following perturbation, blue voxels indicate portions of the vasculature removed by perturbation. Bottom: QuanTAV response scores were re-computed using the vessel network at various levels of perturbation. Robustness of QuanTAV response score was assessed by computing the AUC of the ROC curves at each perturbation level (5, 10, 15, and 20 iterations of perturbation). When compared with the ROC curve of the QuanTAV response score computed with the original skeletons via Delong's test, no level of perturbation was found to produce a significant difference in AUC in either the NSCLC-TRI (p=0.12-0.65) or BRCA-ACT (p=0.11-0.30) cohorts.

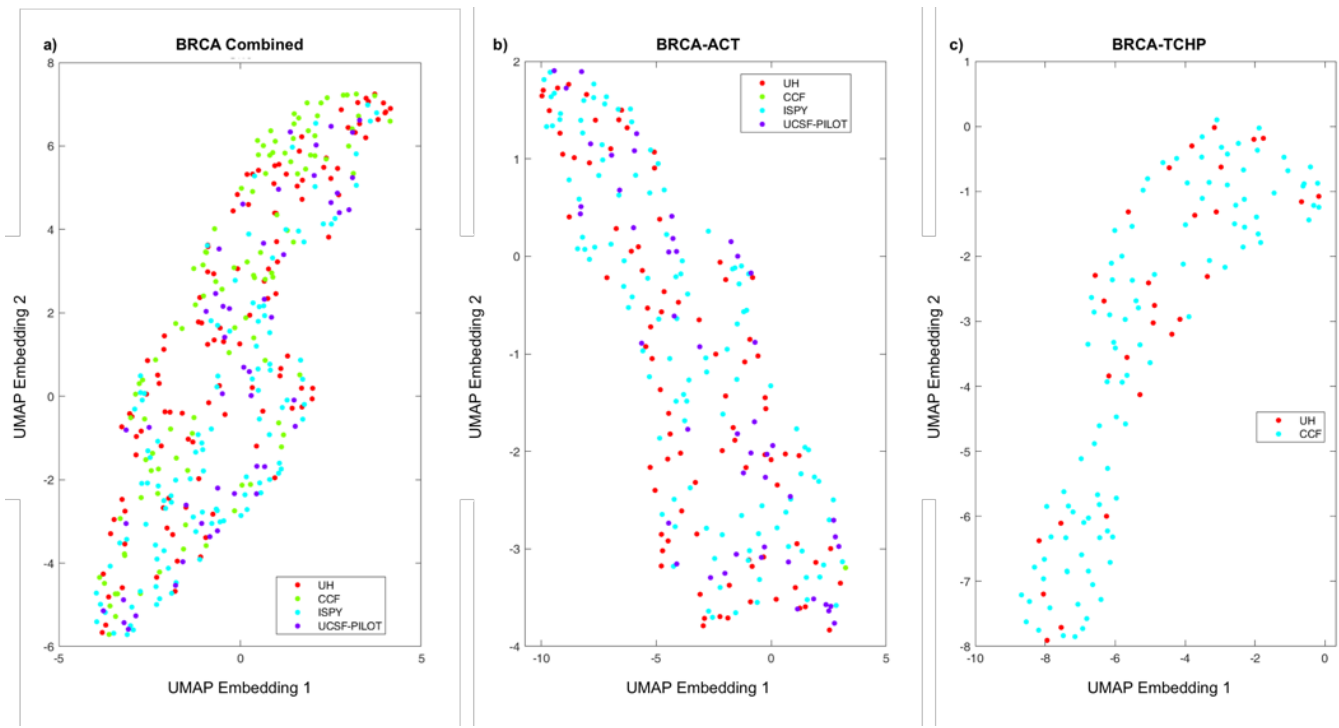

Supplementary Figure 5. UMAP projections of QuanTAV features for (a) all breast cancer patients, (b) the HER2-negative cohort receiving BRCA-ACT, (c) the HER2-positive cohort receiving BRCA-TCHP, shaded according to site. Some separation by site is observed across all patients (a), but this effect disappears when separated by HER2 status/treatment cohort (b&c). UH: University Hospitals, CCF: Cleveland Clinic Foundation, ISPY: ISPY1-TRIAL, UCSF-PILOT: University of California San Francisco ISPY1 pilot study.

# Supplementary Methods – Implementation Details

## Vessel Segmentation

The vesselness filtering utilized to extract the tumor vasculature involves a number of parameters to control the emphasis of vessel-like objects within an image. Supplementary Table 22 includes a description of each parameter and their settings for each modality. Settings were identical between modalities, aside from foreground threshold *C* which was altered on a per image basis for breast MRI scans, due to lack of consistent quantitative values between MRI scanners.[1]

*Supplementary Table 22. Parameters for vesselness filtering of each imaging modality.*

| Parameter | Description | Lung CT value | Breast MRI value |
|---|---|---|---|
| $\alpha$ | Sensitivity parameter for metric that distinguishes between lines (vessel-like objects) and plate-like structures | 0.5 | 0.5 |
| $\beta$ | Sensitivity parameter for metric that distinguishes between lines (vessel-like objects) and blob-like structures | 0.5 | 0.5 |
| C | Threshold for distinguishing background noise and vessel structure. | 20 | ½ the maximum of the Hessian norm* |
| Scale | Parameter specifying expected radius of the detected vessels | 1 | 1 |

*Due to the lack of absolute quantitative intensity values in MRI, *C* was set automatically on a per image basis proportional to the norm of its Hessian matrix. We chose a value of half the maximum of the Hessian norm, a value recommended for images with variable intensity ranges such as MR angiography by Frangi et al.[1]

## QuanTAV Spatial Organization Features

QuanTAV organization features include several tunable parameters that were optimized to each imaging modality and cancer. A grid search was performed to optimize the radius from the tumor to include in spherical vessel projections and the size of the sliding window used to compute local vessel orientations. Vessel distances ranging from 5 to 20 mm from the tumor and sliding window sizes of [20, 35, 50, and 65] pixels were explored in the grid search. Step size for the moving window was fixed to 1/3 of the window size, and spherical coordinates were projected to images of size 400x400 pixels. For each pair of settings evaluated, features were extracted and used to train a classifier in 3-fold cross-validation within the training sets without feature selection. The configuration that maximized the minimum AUC of classifiers across all treatment groups for each modality was chosen as the optimal configuration for that cancer type. Thus, our search prioritized finding a set of QuanTAV organization features that performed well across all treatment contexts for a given imaging modality. Supplementary Figure 6 depicts the performance of QuanTAV organization features at each pair of distance and window settings. The optimal configuration of QuanTAV organization features for breast MRI was found to be a distance of 11 mm from the tumor and a window size of 35 pixels. For NSCLC on CT images, a distance of 7 mm from the tumor and window size of 20 pixels was found to be most effective.

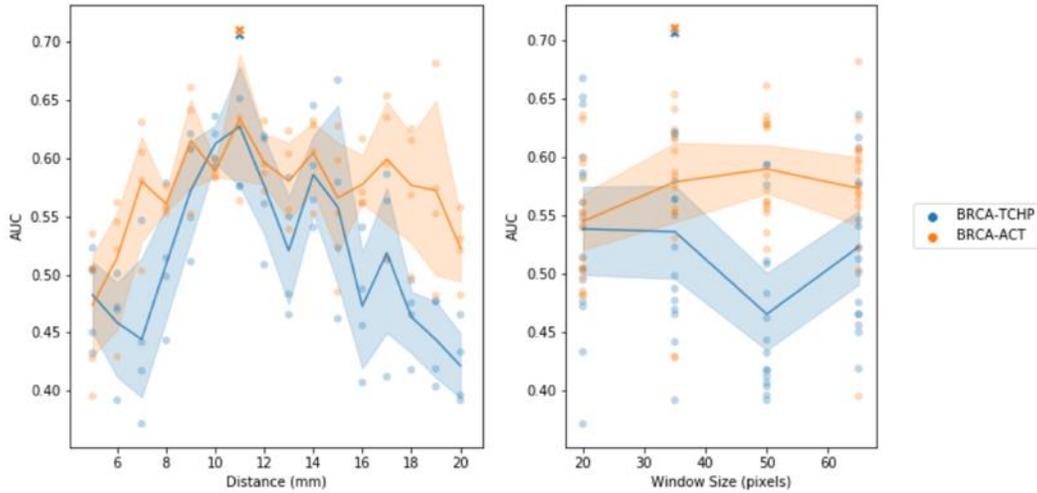

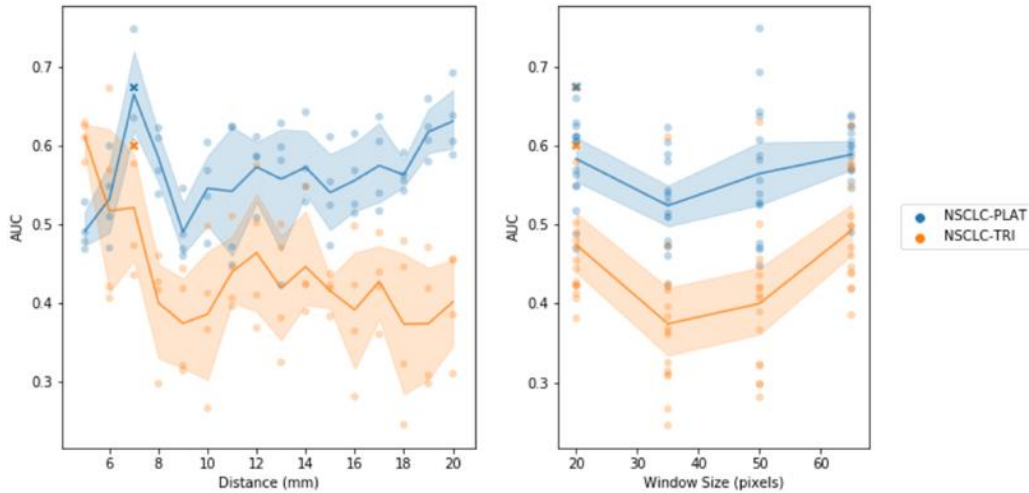

*Supplementary Figure 6. Area under the receiver-operating curve (AUC) in grid searches to optimize QuanTAV Organization features for breast MRI (top), including the BRCA-ACT and BRCA-TCHP treatment groups and chest CT (bottom), including the NSCLC-PLAT and NSCLC-TRI treatment groups. QuanTAV Organization features were extracted at various combinations of inclusion radius and sliding window size settings within each training set and evaluated in cross-validation. For each imaging modality, the settings that maximized the performance of the worst-performing of the two treatment cohort models were chosen as the ideal configuration.*

## *QuanTAV Predictive Response Score*

A two-stage feature selection process was employed to choose a set of QuanTAV features to include in each QuanTAV predictive response score. First, feature selection was performed separately within the two feature groups (QuanTAV Morphology and QuanTAV Spatial Organization) to prune the size of the feature set. A second round of feature selection was then applied to this combined set of remaining QuanTAV Morphology and Spatial Organization features to identify a single best-performing feature set. In both stages, top features were chosen by Wilcoxon rank-sum test in cross-validation. These steps were repeated and re-evaluated with feature sets of sizes between one and six features. The optimal feature selection scheme and size was chosen based on the performance of a linear discriminant analysis (LDA) classifier within the training set in cross-validation.

*QuanTAV Prognostic Risk Score*

The QuanTAV feature set was first reduced to a set of uncorrelated features. The correlation between each pair of features was computed. For a set of correlated features, indicated by a Pearson correlation coefficient (*r*) with an absolute value greater than $r_{min}$, the feature with the greater p value in a two-feature Cox proportional hazards model was removed. From the set of retained, uncorrelated features, a Cox regression model was trained and optimized via 10-fold elastic net regularization with an elastic net mixing parameter of *λ*. The size of the feature set included in each model was determined by the number of features that minimized the deviance in cross-validation, with a minimum and maximum of 5 and 15 features. Values of *λ* and *p* were evaluated in a grid search and chosen based upon maximum performance across all cohorts during nested cross-validation in the training set, with chosen values corresponding to *λ*=0.01 and $r_{min}$=0.8. The coefficient values for the model were then applied to corresponding training and validation sets to derive patient risk scores. Risk score thresholds to stratify patients into high and low risk groups were also derived in the training as previously described[2]. For each prognostic risk score, a cutoff that best separated patients into high and low-risk groups was identified in the training set. Thresholds were first discarded that produced a group smaller than one-quarter of the training set or a log rank test p-value >0.05 were discarded. Among the set of thresholds with the maximum absolute difference in median survival time between groups, the final threshold was chosen by maximum hazard ratio. The derived risk score and risk groups were then applied to the training and testing sets. Models were trained using a modified version of the glmnet package[3] and evaluated using the MatSurv package for survival analysis in MATLAB.[4]

## Supplementary Methods – Additional Experiments

### QuanTAV association with texture-based risk assessment

Within the NSCLC-TRI cohort, a previously published[5] prognostic risk score for trimodality recipients composed of image texture features was assessed for correlations with prognostic QuanTAV features and risk score. The textural risk score consisted of 2 intra-tumoral features and 3 peri-tumoral features extracted within a 15 mm radius from the tumor. The top five most prognostic features of the QuanTAV signature were assessed for correlation with textural risk score in the full NSCLC-TRI cohort. The correlation with overall QuanTAV prognostic risk score was also assessed. In addition, a Cox model combining QuanTAV and texture risk scores was derived in the training set and applied to the testing set to evaluate their potential complementary values in risk assessment.

### Effect of Segmentation Error on QuanTAV signatures.

To investigate the robustness of QuanTAV-based outcome predictions to errors in vessel segmentations, we evaluated the performance of QuanTAV response score at various reduced qualities of vessel segmentation. For each iteration, the set of all branchpoints and endpoints within the vessel skeleton were first identified. At each of these points, the vasculature was randomly perturbed with an equal chance to 1) erode the vessel locally, 2) dilate the vessel locally, or 3) make no change. Degraded vessel segmentations were saved after 5, 10, 15, and 20 iterations of perturbations (depicted in Supplementary Figure 4). Skeletons and QuanTAV features were then re-computed for each perturbed segmentation. The experiment was conducted on the testing sets from one breast (BRCA-ACT) and one lung (NSCLC-TRI) treatment group. QuanTAV response scores were then re-derived on the perturbed testing data and AUC was computed, which was then compared against perform of the original model via DeLong's test of paired ROC curves.[6]

## Supplementary References